\documentclass[aps,prb,twocolumn,showpacs,floatfix]{revtex4}%
\usepackage{amsmath,amsfonts,amssymb}
\usepackage{multirow}
\usepackage{xspace}%
\usepackage[english]{babel}

\usepackage{natbib}

\usepackage{graphicx,color,fancyhdr,xspace}
\definecolor{darkgreen}{rgb}{0,0.5,0}
\definecolor{darkblue}{rgb}{0,0,0.2}
\definecolor{purple}{rgb}{0.35,0,0.35}
\definecolor{orange}{rgb}{1,0.5,0}

\RequirePackage[
   hyperindex,colorlinks,bookmarksnumbered,
   plainpages=true,pdfstartview=FitH]{hyperref}
\hypersetup{linkcolor=blue,urlcolor=darkblue,citecolor=darkgreen}
\usepackage[pdfstartview=FitH]{hyperref}
  \definecolor{checkcolor}{rgb}{1,.2,0}

  \definecolor{wbcolor}{rgb}{0,.2,1}

  \newcommand{\Eq}[1]{Eq.~(\ref{#1})}
  \newcommand{\Eqr}[2]{Eqs.~(\ref{#1}-\ref{#2})}
  \newcommand{\Eqs}[2]{Eqs.~(\ref{#1}) and (\ref{#2})} % PRB style
  \newcommand{\Eqt}[1]{Eq.~\ref{#1}}

  \newcommand{\Fig}[1]{Fig.~\ref{#1}}
  \newcommand{\Figs}[2]{Figs.~\ref{#1} and \ref{#2}} % PRB style

  \newcommand{\ie}{\emph{i.e.}\xspace}
  \newcommand{\eg}{\emph{e.g.}\xspace}
  \newcommand{\vs}{\emph{vs.}\xspace}
  \newcommand{\wrt}{\emph{w.r.t.}\xspace}
  \newcommand{\cf}{\emph{cf.}\xspace}

  \newcommand{\K}{\mathrm{K}}
  \newcommand{\D}{\mathrm{D}}
  \newcommand{\X}{\mathrm{X}}

  \newcommand{\dn}{\ensuremath{n_{0}}\xspace}
  \newcommand{\rhon}{\ensuremath{\hat{\rho}^{[n]}_{0}}\xspace}
  \newcommand{\rhondn}{\ensuremath{\hat{\rho}^{[n;\dn]}_{0}}\xspace}
  \newcommand{\rhosdn}{\ensuremath{\rho^{[n;\dn]}_{s}}\xspace}
  \newcommand{\rhordn}{\ensuremath{\rho^{[n;\dn]}_{r}}\xspace}
  \newcommand{\rhoG}[1]{\ensuremath{\hat{\rho}_{0,#1}}\xspace}

  \newcommand{\Erdn}{\ensuremath{E^{[n;\dn]}_{r}}\xspace}

  \newcommand{\epsXn}{\ensuremath{\varepsilon_{n;\dn}^{\K}\xspace}}
  \newcommand{\epsKn}{\ensuremath{\varepsilon_{n;\dn}^{\K_\chi}}\xspace}
  \newcommand{\epsKx}{\ensuremath{\varepsilon_{n;\dn}^{\K_{(\chi)}}}\xspace}
  \newcommand{\epsDn}{\ensuremath{\varepsilon^{\D}_{n}}\xspace}
  \newcommand{\epsDc}{\ensuremath{\varepsilon^{\D_{\chi}}}\xspace}
  \newcommand{\epsDx}{\ensuremath{\varepsilon^{\D_{(\chi)}}}\xspace}
  \newcommand{\epsDX}{\ensuremath{\varepsilon^{\D}_{(\chi)}}\xspace}
  \newcommand{\epsD}{\ensuremath{\varepsilon^{\D}}\xspace}

  \newcommand{\snr}{\ensuremath{\vert s_{n}\rangle}\xspace}
  \newcommand{\snKl}{\ensuremath{\langle s_{n}^{\K}\vert}\xspace}
  \newcommand{\snKr}{\ensuremath{\vert s_{n}^{\K}\rangle}\xspace}

  \newcommand{\Esn}{\ensuremath{E_{s}^{n}}\xspace}

  \newcommand{\nmin}{\ensuremath{n_{\mathrm{min}}}\xspace}
  \newcommand{\nmax}{\ensuremath{n_{\mathrm{max}}}\xspace}

  \newcommand{\Nkept}{\ensuremath{M_{\K}}\xspace}
  \newcommand{\Ekept}{\ensuremath{E_{\K}}\xspace}
  \newcommand{\trace}[1]{\ensuremath{\mathrm{tr}\left(#1\right)}}

  \newcommand{\nK}{\ensuremath{n_{\mathrm{K}}}\xspace}
  \newcommand{\nB}{\ensuremath{n_{\mathrm{B}}}\xspace}
  \newcommand{\nT}{\ensuremath{n_{\mathrm{T}}}\xspace}
  \newcommand{\TK}{\ensuremath{T_{\mathrm{K}}}\xspace}

\begin{document}

\title{ Discarded weight and entanglement spectra in the Numerical Renormalization Group }
\author{A. \surname{Weichselbaum}}
\affiliation{Physics Department, Arnold Sommerfeld Center for Theoretical Physics, and
Center for NanoScience, Ludwig-Maximilians-Universit\"at, 80333 Munich,
Germany }

\begin{abstract}
A quantitative criterion to prove and analyze convergence within the 
numerical renormalization group (NRG) is introduced. By tracing out a 
few further NRG shells, the resulting reduced density matrices carry 
relevant information on numerical accuracy as well as entanglement. 
Their spectra can be analyzed twofold. The smallest eigenvalues 
provide a sensitive estimate of how much weight is discarded in the 
low energy description of latter iterations. As such, the discarded 
weight indicates in a site-specific manner whether sufficiently many 
states have been kept in a single NRG run. The largest eigenvalues of 
the reduced density matrices, on the other hand, lend themselves to a 
straightforward analysis in terms of entanglement spectra, which can 
be combined into entanglement flow diagrams. The latter show strong 
similarities with the well-known standard energy flow diagram of the 
NRG, supporting the prevalent usage of entanglement spectra to 
characterize different physical regimes. 
\end{abstract}

\date{\today}

% 71.27.+a,   % Strongly correlated electron systems; heavy fermions (Electronic structure of bulk materials)
% 72.15.Qm,   % Scattering mechanisms and Kondo effect (Electronic transport in condensed matter)
% 73.21.La,   % Quantum dots (Electronic structure and electrical properties of ... low-dimensional structures)
% 75.40.Mg,   % Numerical simulation studies (Magnetic properties and materials)

\pacs{
  02.70.-c,   % Computational techniques; simulations (Mathematical methods in physics)
  05.10.Cc,   % Renormalization group methods (Statistical physics, thermodynamics)
  75.20.Hr,   % Local moment in compounds and alloys; Kondo effect (Magnetic properties and materials)
  78.20.Bh    % Theory, models, and numerical simulations (Optical properties, condensed-matter spectroscopy)
}

\maketitle

%% \begin{center}
%%     \hrule\vspace{0.25in}
%%     \begin{minipage}[t]{0.92\linewidth}
%%        \def\tocname{Preliminary table of Contents}
%%        \tableofcontents\vspace{0.25in}
%%     \end{minipage}\hrule
%% \end{center}

\section{Introduction}

The numerical renormalization group (NRG)\cite{Wilson75} is a 
powerful method that provides a highly systematic non-perturbative 
approach to the wide realm of so-called quantum impurity systems. 
These consist of an arbitrary small quantum system (the 
\emph{impurity}) in contact with a macroscopic non-interacting 
usually fermionic bath. Each part is simple to solve exactly on its 
own. In presence of interaction at the location of the impurity, 
however, the combination of both gives rise to strongly-correlated 
quantum-many-body phenomena. \cite{Bulla08} 
Wilson's logarithmic coarse-graining of the bath leads to a 
semi-infinite chain with exponentially decaying couplings, which 
justifies the concept of energy scale separation. That is the Wilson 
chain can be diagonalized iteratively by adding one site at a time 
and retaining the lowest \Nkept states only. The obvious question, 
however, is how many states should one keep on average for 
convergence in this procedure? At a given iteration there is no 
quantitative \emph{a priori} measure that indicates how many 
low-energy states are required for a proper description of the 
remaining low-energy physics. Usually, the only way to check 
convergence within the NRG is by repeating the entire calculation and 
showing that the results no longer change when further increasing 
\Nkept. Therefore an NRG calculation is typically run somewhat 
blindly for some pre-determined \Nkept. 

\begin{figure}[b]
\begin{center}
\includegraphics[width=.9\linewidth]{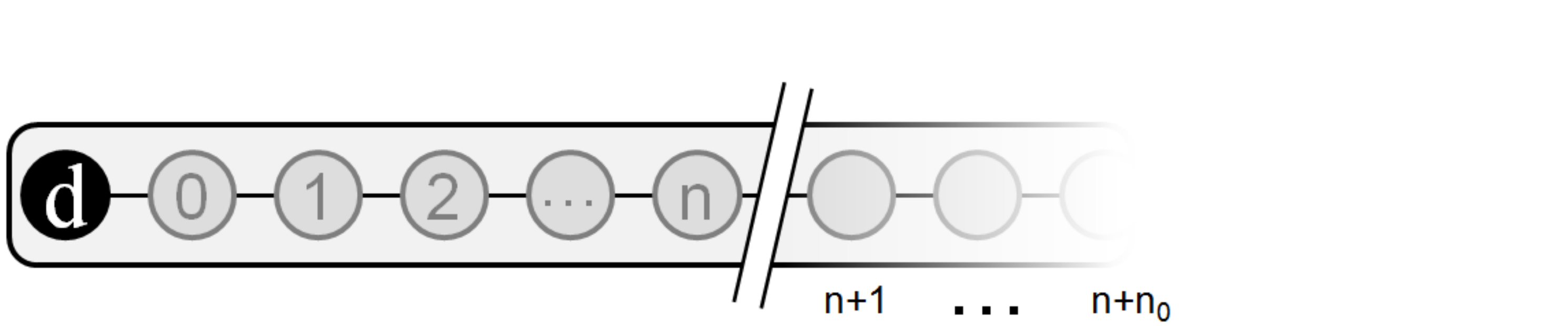}
\end{center}
\caption{ Schematic depiction of tracing out the low-energy
sector of the Wilson chain at iteration $n$ by including 
and analyzing \dn more NRG iterations. The impurity (dot)
is entirely contained in the first site, while the bath
is coarse-grained and mapped onto the remaining semi-infinite
tight-binding chain of sites $n=0,1,2,\ldots$.}%
\label{fig:wilson-chain}%
\end{figure}

This somewhat uncontrolled truncation in the NRG is in stark contrast 
to the situation in the density matrix renormalization group (DMRG). 
\cite{White92,Schollwoeck05,Schollwoeck11} DMRG is based on a 
(strictly) variational principle, and as such has a clean 
well-defined truncation of the state space for part of the system 
through the discarded weight in its reduced density matrix. 
\cite{Schollwoeck05} In contrast to the less suggestive plain number 
\Nkept of states kept, the discarded weight represents a reliable 
\emph{quantitative} measure for the accuracy of a calculation. Within 
the DMRG, \Nkept can be easily adjusted according to some predefined 
threshold in the discarded weight, instead. 
Motivated by DMRG then, an approximate similar criterion can be 
established within the NRG as will be shown in the following. The 
analysis requires a slightly longer chain, as shown schematically in 
\Fig{fig:wilson-chain}. With the extra \dn sites traced out again 
from the ground state space of the enlarged system, this allows to 
estimate the discarded weight. The latter offers a quantitative 
convergence measure that is specifically of interest for numerically 
expensive models such as multi-channel models, or models where the 
energy scale separation along the Wilson chain might be in question 
due to modifications in the discretized Hamiltonian. In either case, 
a small discarded weight provides a strong indication for converged 
NRG data. 

Furthermore, the reduced density matrices generated for the 
evaluation of the discarded weight also allow a quite different 
analysis in terms of their \emph{dominant} correlations. In 
particular, combining their entanglement spectra into 
\emph{entanglement flow diagrams}, this offers a complementary view 
to the usual NRG energy flow diagram, which is entirely based on the 
analysis of the low-energy state space of a prior NRG run. 

The paper is thus organized as follows. In Sec. I the essentials of 
the numerical renormalization group are revisited, including the 
construction of reduced density matrices. Sec. II then uses a 
specific set of reduced density matrices in the definition and 
analysis of the discarded weight within the NRG. Sec. III offers a 
complementary view on these reduced density matrices by analyzing 
their entanglement content in terms of entanglement spectra. Sec. IV, 
finally, summarizes and includes an outlook.

\subsection{The numerical renormalization group}

Within the NRG, the continuum of the non-interacting bath of 
half-bandwidth $W$ is logarithmically coarse-grained in energy space, 
followed by an exact mapping onto a semi-infinite so-called 
Wilson-chain. \cite{Wilson75,Krishna80I} The impurity space is 
coupled to the first site of this chain only, as depicted 
schematically in \Fig{fig:wilson-chain}. The logarithmic 
coarse-graining is defined through the dimensionless discretization 
parameter $\Lambda>1$. With the chemical potential at energy zero, 
the continuum of states in the energy intervals $\pm 
W[\Lambda^{-(n-z+1)}, \Lambda^{-(n-z)}]$ is effectively represented 
by single fermionic levels (coarse-graining), including an arbitrary 
z-shift with $z\in[0,1[$. \cite{Yoshida90,Oliveira92,Zitko09} The 
subsequent exact mapping onto the semi-infinite chain (Lanczos 
tridiagonalization) \cite{Demmel00} results in an effective 
tight-binding chain with the \textit{exponentially} decaying hopping 
$t_{n} \sim\Lambda^{-n/2}$ between sites $n$ and $n+1$. For 
sufficiently large $\Lambda$, typically $\Lambda\gtrsim 1.7$, this 
then justifies the essential NRG assumption of \textit{energy scale 
separation}: by iterative diagonalization of the Wilson chain by 
adding one site at a time, large energies are considered first, with 
the (approximate) eigenstates at large energies \textit{discarded} 
and considered unimportant in the description of the lower energy 
scales still to follow. Thus each site of the Wilson chain 
corresponds to an energy shell with a characteristic energy scale, 
\begin{align}
  \omega_{n} \equiv \tfrac{\Lambda^{z-1}(\Lambda-1)}{\log \Lambda}
  W\Lambda^{-\tfrac{n}{2}} \text{.}
\label{eq:def-escale}
\end{align} 
Here, the prefactor was chosen such, that the rescaled couplings 
$\lim_{n\to\infty}\left(t_n/\omega_{n}\right)=1$ quickly approach 
unity for longer Wilson chains for arbitrary $\Lambda$ and z-shift, 
with the discretization following the prescription of 
[\onlinecite{Zitko09}] for a flat hybridization, \ie 
$\Gamma(E)=\Gamma \theta(W-|E|)$. 

With $\hat{H}_{n}$ the full Hamiltonian $\hat{H}$ of the Wilson chain 
up to and including site $n$, its low-energy eigenstates are given by 
the NRG eigenstates $\hat{H}_{n} \vert s_{n}\rangle = E_{s}^{n} \vert 
s_{n}\rangle$. Complemented by an arbitrary state $\vert e_{n}\rangle 
$ for the remainder of the system following site $n$, the NRG 
assumption of energy scale separation can be summarized then in the 
following approximation, \cite{Anders05} 
\begin{equation}
   \hat{H} \vert se \rangle_{n}
   \simeq E_{s}^{n} \vert se \rangle_{n}
\text{,}\label{eq:NRG-approx}%
\end{equation}
\ie the states $\vert se \rangle_{n} \equiv \vert s_{n}\rangle 
\otimes \vert e_{n}\rangle$ are, to a good approximation, also 
eigenstates of the entire Wilson chain. The energies $E_{s}^{n} $ at 
iteration $n$ are usually expressed relative to the ground state 
energy of that iteration, and rescaled by a factor 
$\tfrac{W}{2}\left( \Lambda+1\right) \Lambda^{-n/2} \propto \omega_n$ 
to resolve the energy shell at iteration $n$. The resulting energies 
are referred to as \textit{rescaled energies}. For fully fermionic 
systems, they typically show an intrinsic even-odd behavior. Thus 
combining the rescaled energies \vs even and odd iterations $n$
separately, this results in the standard \emph{energy flow diagrams} 
of the NRG. \cite{Wilson75,Krishna80I}

The approximate many-body eigenstates $\left\vert se\right\rangle 
_{n}$ are constructed iteratively, and therefore described in terms 
of matrix-product-states. \cite{Wb07,Wb09,Rommer95,Schollwoeck11} 
Each iterative step results in a basis transformation, encoded in an 
A-tensor, that combines an existing effective basis $\vert s_{n} 
\rangle$ for the system up to and including site $n$ with the state 
space $\vert \sigma \rangle $ of site $n+1$, 
\begin{equation}
   \vert s_{n+1}\rangle 
 = \sum_{s_{n}^{\prime},\sigma_{n+1}}  \vert s_{n}^{\prime},\sigma_{n+1} \rangle
   \underset{\equiv A_{s_{n}^{\prime} s_{n+1}}^{[\sigma_{n+1}]}}
  {\underbrace{\langle s_{n}^{\prime},\sigma_{n+1}|s_{n+1} \rangle }}
\text{,} \label{eq:Atensor}%
\end{equation}
with $\vert s_{n}^{\prime},\sigma_{n+1} \rangle \equiv \vert 
s^{\prime}_{n}\rangle \otimes \vert \sigma_{n+1}\rangle$. The 
orthogonality of state spaces, $\langle s_{n+1} \vert 
s^{\prime}_{n+1} \rangle = \delta_{ss^{\prime}}$, directly implies 
the orthonormality relation for $A$-tensors,\cite{Schollwoeck05} 
\begin{equation}
\sum_{\sigma_{n+1}}
   A^{\left[ \sigma _{n+1}\right] \dagger}
   A^{\left[ \sigma _{n+1}\right] } = \mathbf{1}
\text{.} \label{eq:Aortho}%
\end{equation}
Without truncation, the dimension $M_{n}\ $of the state space $\vert 
s_{n}\rangle$ increases exponentially with the number of sites 
included, $M_{n}\sim d^{n}$, with $d$ the dimension of a local Wilson 
site. Therefore the maximum number of states \Nkept, that one can 
maintain in a calculation, is quickly reached after $\dn \simeq 
\log(\Nkept) / \log(d)$ iterations. For every subsequent iteration, 
the state space $\vert s_{n}\rangle$ is truncated by retaining the 
lowest \Nkept states in energy only. This leads to the distinction 
between $\vert s_{n}^{\K}\rangle$ and $\vert s_{n}^{\D}\rangle$ for 
kept and discarded states at iteration $n$, respectively. 
Correspondingly, this also splits the A-tensor into two parts, 
$A_{KK}$ and $A_{KD}$, that propagate the state kept space from the 
previous iteration into the newly generated kept or discarded space, 
respectively. 

The truncation criteria with respect to a fixed prespecified \Nkept 
can be softened towards a energy cutoff, \cite{Bulla08} \Ekept, that 
is taken constant in rescaled energies. For a fair comparison for 
different z-shifts, it will be specified in units of the energy scale 
$\omega_{n}$ in \Eq{eq:def-escale}. Since NRG data typically appears 
bunched at certain energies (\eg see \Fig{fig:nrgtruncDd} later), 
\Ekept may hit a ``gap'' in the NRG spectrum at some iteration, and 
the last ``bunch'' of states included may lie, on average, at clearly 
smaller energies than \Ekept. Given the empirical importance of the 
first few NRG iterations, therefore as a safety measure, by default, 
\Ekept was taken by 20\% larger for the very first iteration where 
truncation occurred, \ie using 1.2\Ekept there with \Ekept specified 
in context. Typical values are in the range $\Ekept =5\ldots8$. 

%% \subsubsection{Model system}

The model system considered in this paper is the well-known standard 
single impurity Anderson model (SIAM), 
\begin{align}
    H_{N}^{\mathrm{SIAM}}  &  =
    \sum_{\sigma}\varepsilon_{d\sigma}\hat{n}_{\sigma}
  + U\hat{n}_{d\uparrow}\hat{n}_{d\downarrow}
  + \sum_{\sigma}\sqrt{\tfrac{2\Gamma}{\pi}}
     \bigl(  \hat{d}_{\sigma}^{\dagger}\hat{f}_{0\sigma}+h.c.\bigr) \nonumber \\
& + \sum_{\sigma}\sum_{n=0}^{N-1}t_{n}
    \bigl(  \hat{f}_{n,\sigma}^{\dagger}\hat{f}_{n+1,\sigma}^{{}}+h.c.\bigr) 
\label{eq:H-SIAM}%
\end{align}
with the operators $\hat{d}_{\sigma}^{\dagger}$ ($\hat{f}_{n\sigma 
}^{\dagger}$) creating a particle with spin 
$\sigma\in\{\uparrow,\downarrow\}$ at the impurity (at site $n$ in 
the bath), respectively, having 
$\hat{n}_{d\sigma}\equiv\hat{d}_{\sigma}^{\dagger}\hat {d}_{\sigma}$.
%% and $\hat{n}_{n\sigma}\equiv\hat{f}_{n\sigma}^{\dagger}\hat 
%% {f}_{n\sigma}$. 
The energy $\varepsilon_{d\sigma} \equiv \varepsilon_{d} - 
\tfrac{B}{2}(\hat{n}_{d\uparrow} - \hat{n}_{d\downarrow})$ is the 
spin dependent level-position of the impurity in the presence of a 
magnetic field $B$. Furthermore, $U$ is the onsite Coulomb 
interaction and $\Gamma$ the hybridization of the impurity with the 
bath. All parameters will be specified in units of the bandwidth 
$W:=1$ in context with the figure panels. The bath in \Eq{eq:H-SIAM} 
is already represented in terms of a Wilson chain, \cite{Wilson75} 
described by the semi-infinite tight binding chain ($N\to\infty$) 
with exponentially decaying hopping amplitudes $t_{n} 
\sim\Lambda^{-n/2}$. In practice, $N$ can be taken finite, with 
$\hat{H}_{n}$ describing the Wilson chain up to and including site $n 
\le N$. 

Charge and spin are conserved in the SIAM in \Eq{eq:H-SIAM}, where, 
however, only the abelian part of the symmetries is included in the 
calculations. Hence the number of states \Nkept directly refers to 
the actual number of states kept in a calculation (in contrast to the 
dimension of reduced multiplet spaces with non-abelian symmetries). 
Similarly, also the discussion of the entanglement spectra further 
below will refer to the abelian symmetry labels which also applies 
when non-abelian symmetries are broken. Note that while, in general, 
a particle-hole symmetric impurity setting will be used, this can be 
easily broken by applying a (small) gating potential to the impurity 
level. Moreover, the SU(2) spin symmetry, in fact, will be broken 
explicitly by the application of an external magnetic field. 

\subsection{Density matrices}

The NRG eigenbasis of \Eq{eq:NRG-approx} with respect to the 
discarded space forms a complete many-body eigenbasis. 
\cite{Anders05} Initially introduced for the feat of real 
time-evolution within the NRG, this eigenbasis is actually applicable 
and tractable more generally within the NRG framework. 
\cite{Peters06} In particular, this allows the clean calculation of 
correlation functions in terms of the full density matrix (FDM) in 
the many-body eigenbasis, \cite{Wb07} in that 
\begin{align}
   \hat{\rho}(T) &\equiv \tfrac{1}{Z} e^{-\beta\hat{H} } \cong
   \tfrac{1}{Z} \sum_{nse} e^{-\beta E_{s}^{n}}
   \vert se \rangle_{n\,n\!}^{\D\,\!\D\!} \langle se \vert
\text{,}\label{eq:FDM-rho}
\end{align}
with $\beta\equiv1/k_{B}T$ for arbitrary temperatures $T$, using 
\emph{non-rescaled} energies $E_{s}^{n}$ relative to a common energy 
reference, by construction of a thermal density matrix. 
\Eq{eq:FDM-rho} can be rewritten as $\hat{\rho}(T)\equiv \sum_n 
w_n(T) \hat{\rho}_n(T)$, \ie a normalized distribution $\sum_n w_n=1$ 
of the density matrices $\hat{\rho}_n(T)$ generated in the basis of 
iteration $n$. \cite{Wb07} For a given temperature $T$, the 
distribution $w_n$ is strongly peaked around iteration \nT that 
corresponds to the energy scale of temperature. Hence temperature 
essentially terminates the Wilson chain.

\begin{figure}[t]
\begin{center}
\includegraphics[width=.55\linewidth]{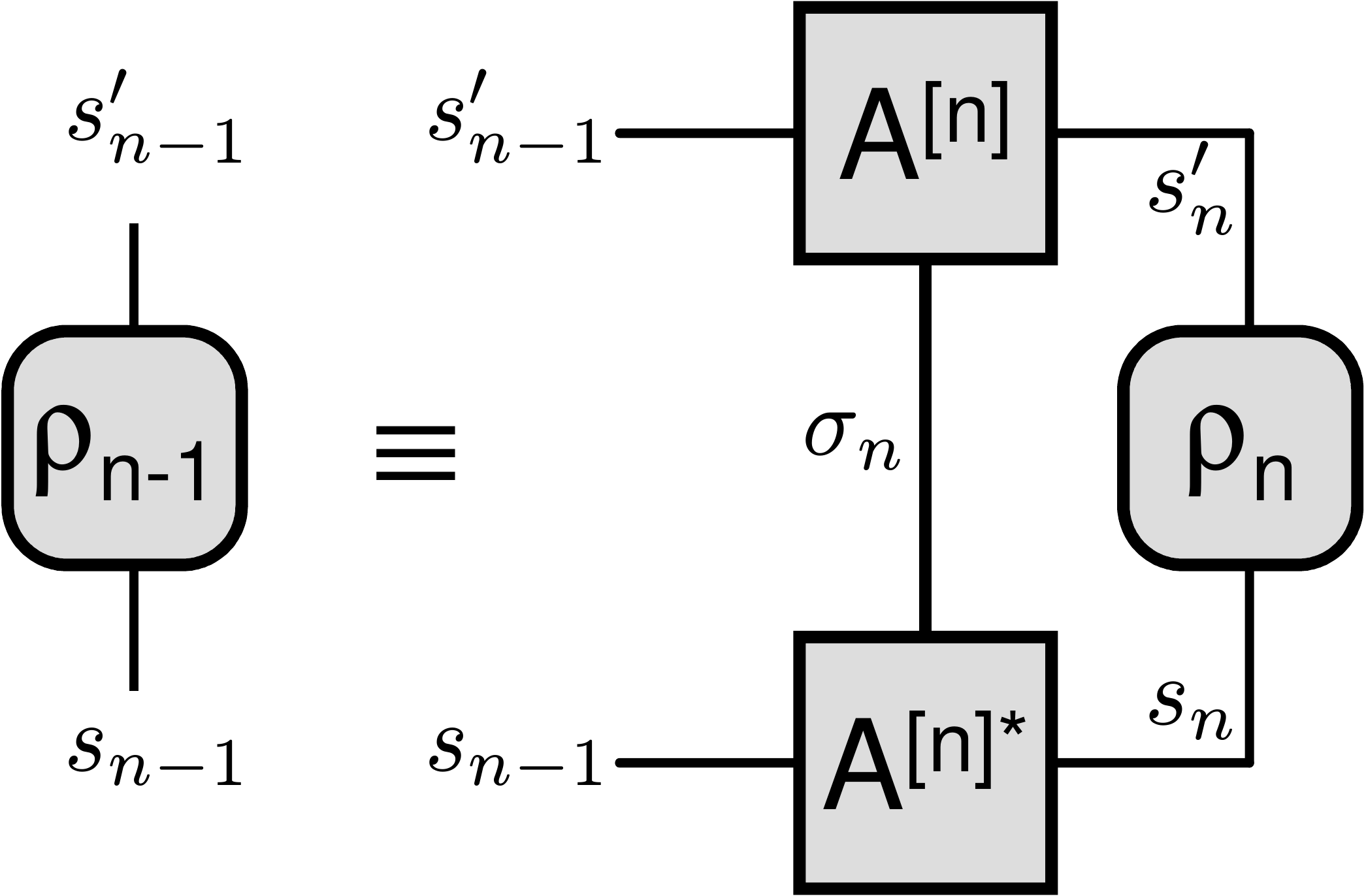}
\end{center}
\caption{ \emph{Backward update} of given density matrix $\rho_{n}$ at
iteration $n$. Blocks represent data spaces, lines correspond to indices.
The lines connecting different blocks are contracted indices (\ie indices
summed over), such as $\sigma_{n}$, $s_{n}$ and $s_{n}^{\prime}$, while open
lines represent open indices, \eg the indices $s_{n-1}$ and $s_{n-1}^{\prime
}$.}%
\label{fig:rhoupdate}%
\end{figure}

In this paper, however, mainly reduced density matrices derived from 
ground states will be considered, hence temperature is essentially 
zero. More generally then, consider an arbitrary density matrix 
defined in the many-body basis $\vert s_{n} \rangle$ of iteration $n$ 
in either kept or discarded space, $\X\in \{ \K,\D\} $, 
\begin{align}
   \hat{\rho}_{n}^{\left[ \X\right] } \equiv \sum_{s_{n}s_{n}^{\prime} \in \X}
   \rho_{s_{n}s_{n}^{\prime}}^{[\X]}
   \vert s_{n} \rangle \langle s_{n}^{\prime} \vert
\text{,}
\end{align}
where $\rho_{n}^{\left[ \X\right] }$ (\ie without the hat) represents 
the space of matrix elements $\rho_{s_{n}s_{n}^{\prime}}^{[\X]}$. The 
prototypical and well-known operation on such a density matrix is 
tracing out the last site $n$, 
\cite{Hofstetter00,Anders05,Peters06,Wb07,Toth08} 
\begin{align}
   \hat{\rho}_{n-1}^{[\K]}  & =
   \sum_{\substack{s_{n-1},s_{n-1}^{\prime }\\\sigma_{n}}}
   \Bigl(  A_{KX}^{[\sigma_{n}]} \rho^{[\X]}_n A_{KX}^{[\sigma_{n}]\dagger}
   \Bigr)_{s_{n-1} s_{n-1}^{\prime}}
   \vert s_{n-1}\rangle \langle s_{n-1}^{\prime}\vert
   \nonumber\\
& \equiv\mathcal{\hat{P}}_{n}\hat{\rho}_{n}^{[\X]}\text{,}
\label{eq:backprop}%
\end{align}
written as a matrix product of the matrices 
$A_{KX}^{[\sigma_{n}](\dagger)}$ and $\rho^{[\X]}_n$ in the first 
line. Equation~(\ref{eq:backprop}), in the following referred to as 
\textit{backward} \textit{update}, introduces the notational 
shorthand $\mathcal{\hat{P}}_{n}$ for the bilinear product of the 
$A$- and $A^{\ast}$-tensor at site $n$, that acts as a linear 
superoperator on the density matrix $\hat{\rho}_{n}$. The 
corresponding contraction pattern is shown in a simple graphical 
depiction in \Fig{fig:rhoupdate}. By construction, the backward 
update of a density matrix in \Eq{eq:backprop} always results in a 
density matrix in the kept space of the earlier iteration, and with 
\Eq{eq:Aortho} representing a complete positive map, \Eq{eq:backprop} 
clearly also preserves the properties of a density matrix. 

\section{Discarded weight within the NRG}

The standard notion of NRG is that it zooms in towards the low energy 
sector of a given many-body Hamiltonian, while iteratively discarding 
states at higher energies. Having a semi-infinite chain, this can 
continue to arbitrarily small energy scales, which enables NRG to 
resolve dynamically generated small energy scales as they appear, for 
example, in the context of Kondo physics. From a variational point of 
view for matrix-product-states, this implies that the cost function 
can be identified as 
\begin{align}
   \lim_{N\to\infty} \langle s_N| H_N | s_N\rangle
   &\to \mathrm{MIN}
\text{,}\label{eq:NRGcost}
\end{align}
yielding the ground state $|0\rangle_\infty$ of the semi-infinite 
Wilson chain. For a sufficiently long chain of total length $N$ then 
included in a given calculation, the state $|0\rangle_N$ will be 
referred to as the overall ground state of this Wilson chain. In 
fact, the cost function in \Eq{eq:NRGcost} is well captured within 
the NRG through its principle of energy scale separation. 
\cite{Saberi08} 

If at a given iteration within the NRG states essentially decouple 
with respect to the low energy state space still to follow, these 
states will quickly and efficiently be discarded as high energy 
states. The truncation towards the low-energy sector also implies, 
that the state space at large energies is necessarily more crudely 
resolved, consistent with the coarser discretization there. The 
lowest \Nkept states kept at a given iteration $n$ then are important 
for the correct description of the low-energy sector still to come. 
However, there is no real quantitative \emph{a-priori} measure to 
indicate whether the number \Nkept of states to be kept is 
appropriate. Conversely, however, at a given iteration $n$ one can 
ask whether all states kept a few iterations earlier were actually 
important. This question can be answered entirely within the kept 
spaces of these iterations, hence is numerically cheap to analyze. 

\subsection{Construction of reduced density matrices}

Consider the actual ground state space $G$ at some arbitrary but 
fixed iteration $n'$. In general, it may be $g_{n'}$-fold degenerate, 
hence consider its fully mixed density matrix, 
\begin{align}
   \rhoG{n'} 
 & \equiv \tfrac{1}{g_{n'}} \sum_{s\in G} \vert s_{n'} \rangle 
   \langle s_{n'} \vert \text{.}
\end{align}
By construction, the number of eigenvalues of \rhoG{n'} unequal zero, 
\ie its Schmidt rank, is equal to $g_{n'}$. Now, tracing out the last 
iteration $n'$, \ie the lowest energy scale included in \rhoG{n'}, is 
equivalent to the back-propagation $\hat{\rho} ^{[n'-1;1]}_{0} \equiv 
\mathcal{\hat{P}}_{n'} \rhoG{n'}$ in \Eq{eq:backprop}. Through this 
operation, the Schmidt rank will rise, in general, by a factor of 
$d$, with $d$ the state space dimension of a Wilson site. Repeating 
this process iteratively, this allows to trace out the $\dn$ smallest 
energy shells in \rhoG{n'}. Thus with $n'=n+\dn$, this leads to the 
reduced density matrix, 
\begin{align}
   \rhondn &  \equiv
   \bigl(\prod_{l=n+1}%
   ^{n+\dn} \mathcal{\hat{P}}_l\bigr) \rhoG{n+\dn} \nonumber\\
&  \equiv\sum_{ss^{\prime} }^{\Nkept}\rho_{ss^{\prime}}^{[n;\dn]}
   \vert s_{n}^{\K} \rangle \langle s^{\prime\K}_{n} \vert
\text{,} \label{eq:rho_nn0}
\end{align}
which, by construction, is defined in the \textit{kept} space of 
iteration $n$. The Schmidt rank will grow quickly, \ie exponentially, 
in this process, until after $\dn$ iterations, with 
\begin{equation}
   \dn \gtrsim \mathrm{ceil}\bigl[\log(\Nkept)/\log(d)\bigr]\qquad (\dn\ll N)
\text{,}
\label{eq:def_n0}%
\end{equation}
it reaches the full dimension \Nkept of the kept space. Typically, 
$\dn$ is much smaller compared to the full length $N$ of the Wilson 
chain considered, and conversely also specifies the initial number of 
NRG iterations in a forward direction that can be typically performed 
without truncation. For the definition of the discarded weight below, 
it is sufficient to stop the back-propagation of \rhoG{n+\dn} at this 
point. 

The reduced density matrix \rhondn generated in \Eq{eq:rho_nn0} is, 
in general, not diagonal in the energy eigenbasis $\vert s_{n}^{\K} 
\rangle$, since through the traced out lower-energy sites it does 
know about an enlarged system. Its eigenvectors are described by a 
unitary transformation $u_{rs'}^{[n;\dn]}$ within the NRG eigenstates 
kept at iteration $n$, 
\begin{align} 
   &|r_{n;\dn}\rangle
   \equiv \sum_{s'} u_{rs'}^{[n;\dn]} |s_n^{\prime\K}\rangle, \nonumber\\
  \text{with}\quad &\rhondn |r_{n;\dn}\rangle
 = \rho_r^{[n;\dn]} |r_{n;\dn} \rangle
\text{,}\label{eq:rho-eig}
\end{align}
where the index $r$ shall refer to the eigenstates of the reduced 
density matrix, in contrast to the index $s$ for the energy 
eigenstates. Here, the eigenvalue \rhordn describes the importance of 
a specific linear superposition of NRG eigenstates at iteration $n$ 
for the low-energy description of latter iterations. 

This offers two routes for the analysis of the density matrices 
\rhondn. (i) Adhering to the energy eigenbasis of the NRG, the 
importance of the kept state \snKr at eigenenergy \Esn for the latter 
low-energy physics is given by the expectation value
\begin{align}
  \rhosdn \equiv \snKl \rhondn \snKr
\text{,}\label{eq:rdiag}
\end{align}
\ie the diagonal matrix elements $\rho_{ss}^{[n;\dn]}$. 
Alternatively, (ii) using the eigenbasis of the reduced density 
matrices, the weights of these states are given by the eigenvalues 
\rhordn, while now their energies are given by the expectation values 
\begin{align}
  \Erdn \equiv \langle r_{n;\dn}| \hat{H}_n |r_{n;\dn}\rangle
\text{.}\label{eq:reig}
\end{align}
Both routes will be analyzed and compared in the following. However, 
the actual eigendecomposition of the reduced density matrices will be 
preferred for the remainder of the paper as explained. 

In either case, a set of states $i$ with (average) energy $E_i$ is 
given together with their respective (average) weight $\rho_i$ that 
represents the states importance for latter iterations. For the first 
[second] route above this data is given by $(\Esn,\rhosdn)$ 
[$(\Erdn,\rhordn)$], respectively. Given that the reduced density 
matrix \rhondn, by construction, exists in the kept space only, 
therefore also all states $i$ refer to the kept space or a linear 
superpositions thereof. Moreover, for every iteration, the weights 
$\rho_i$ are normalized, \ie they are positive and add up to 1, while 
by combining data from different iterations, the energies $E_i$ are 
always specified in rescaled units. 

\begin{figure}[tb!]
\begin{center}
\includegraphics[width=1\linewidth]{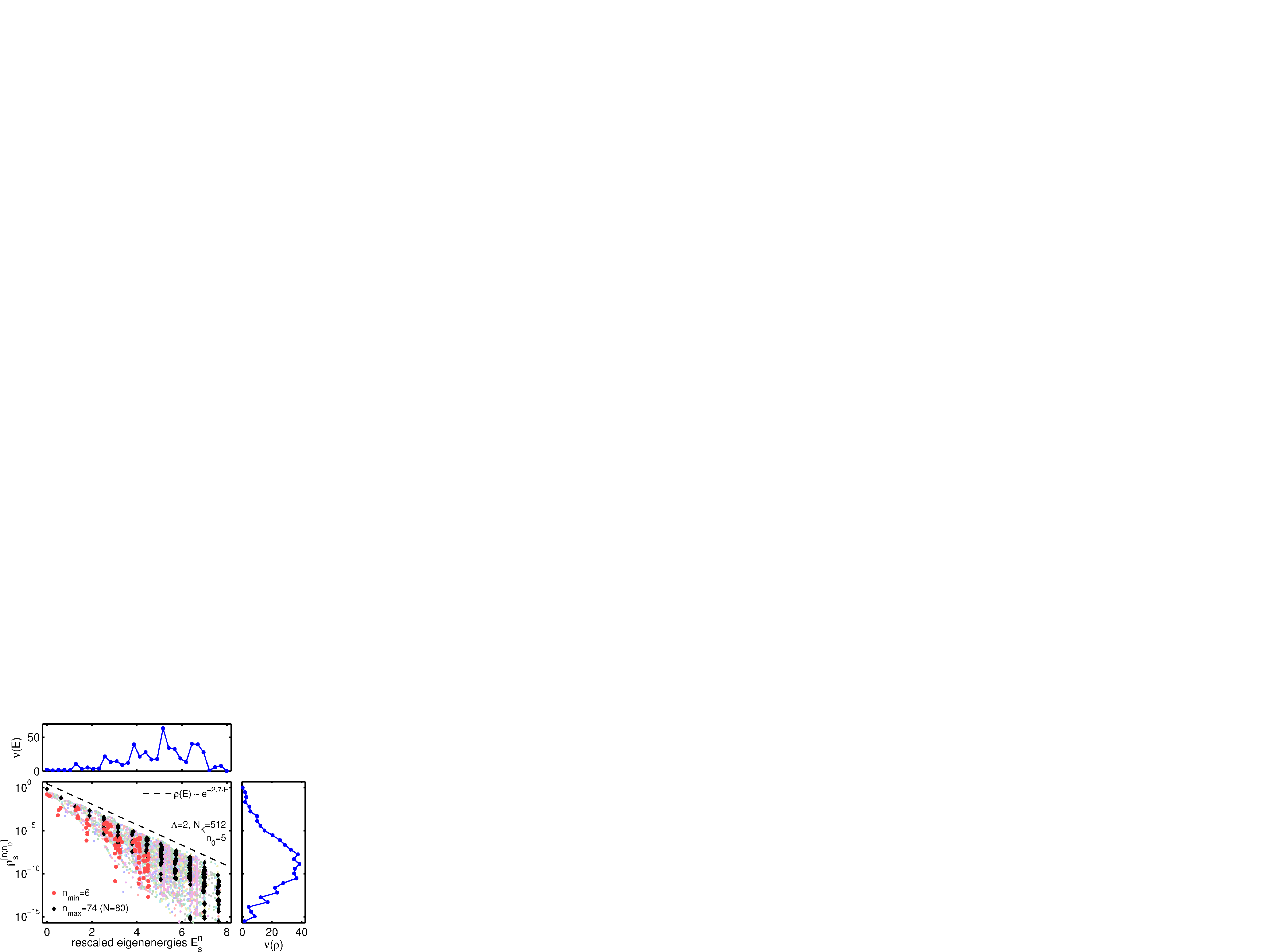}
\end{center}
\caption{ (Color online) %
Weight distribution of energy eigenstates over full NRG run at fixed 
$\Nkept=512$ for the SIAM [\Eq{eq:H-SIAM}: $U=0.20, ~\epsilon_{d} = 
-U/2, ~\Gamma =0.01$]. The main panel shows the rescaled eigenenergies
\Esn vs. their weights \rhosdn as in \Eq{eq:rdiag}.
Data is shown only for those iterations where truncation occurred, 
with data from the same iteration shown in the same color. The two 
iterations with smallest (largest) energy range, \nmax (\nmin), are 
highlighted in strong colors (black diamonds (red crosses)), 
respectively, while light colors are used for all other iterations.
The top [right] panel shows the energy [weight] distribution 
$\nu(E)$ [$\nu(\rho)$], \Eq{nu:E} [\Eq{nu:rho}], respectively, for the 
data in the main panel, with matching energy $E$ [weight $\rho$] axis.
The binning referred to in the text to \Eqs{nu:E}{nu:rho} is indicated
by the intervals between the data points in the top and right panel.
}% 
\label{fig:nrgtruncDd} 
\end{figure}

The resulting data $(E_i,\rho_i)$ then is clearly correlated. It is 
analyzed threefold, (i) in terms of the average distribution of the 
rescaled energies $E_i$ 
\begin{equation}
   \nu(E)\cong\tfrac{1}{N'}\Bigl.
   \sum_{n=1}^{N}\Bigr.^\prime \sum_{E<E_{i}<E+dE} 1
\text{,}\label{nu:E}%
\end{equation}
\label{nu:rho+E}%
(ii) the average distribution of the weights $\rho_i$, 
\begin{equation}
   \nu(\rho)\cong\tfrac{1}{N'}\Bigl.
   \sum_{n=1}^{N}\Bigr.^\prime \sum_{\rho<\rho_{i}<\rho+d\rho} 1
\text{,}\label{nu:rho}%
\end{equation}
and (iii) their average dependence on each other 
\begin{subequations}
\begin{align}
   \rho(E)  &\cong \tfrac{1}{N'dE}
   \sum_{n}^{N}\Bigr.^\prime \sum_{E<E_{i} <E+dE} \rho_i
\label{eq:def-rhoE} \\
   &\cong \kappa e^{-\kappa E}
\text{.}\label{eq:rhoE-decay}%
\end{align}
\label{eq:rhoE}
\end{subequations}
Here some appropriate linear (logarithmic) binning of the data is 
assumed with energy (weight) intervals $dE$ ($d\rho$), respectively. 
In particular, the densities in \Eqs{nu:E}{nu:rho} are 
clearly dependent on these binning intervals, which therefore will 
properly indicated in the subsequent plots. The prime in the 
summation and the normalization indicates that only those iterations 
$n$ are included where state space truncation occurred, \ie typically 
$n \gtrsim \dn$. The total number of these iterations is given by 
$N'$. With chosen normalization then, the sum over the binned 
$\nu(E)$ and $\nu(\rho)$ data both yield the average number of kept 
states, while the integrated weight distribution $\rho(E)$ in 
\Eq{eq:def-rhoE} is normalized to 1, since $\trace{\rho} \sim 
\int_0^\infty \rho(E)\,dE = 1$. As will be seen later, the weight 
distribution $\rho(E)$ typically shows a clear exponential decay with 
a characteristic exponent $\kappa$, as indicated already in 
\Eq{eq:rhoE-decay}, with the prefactor chosen such that it also 
preserves normalization. 

\subsubsection{Energy eigenbasis}

The correlation between the eigenenergies \Esn and their 
corresponding weights \rhosdn is plotted as a scatter plot in the 
main panel of \Fig{fig:nrgtruncDd}. The model analyzed is the SIAM in 
\Eq{eq:H-SIAM} in the Kondo regime using a fixed number of kept 
states, with all parameters specified in the figure caption. The 
weights \rhosdn clearly diminish exponentially with energy, which is 
intuitively expected as a consequence of energy scale separation 
within the NRG. The integrated weight distribution $\rho(E)$ (dashed 
black line, \cf \Eqt{eq:def-rhoE}), shows a clear exponential decay 
with an exponent $\kappa\simeq 2.7$. As seen in 
\Fig{fig:nrgtruncDd}, this distribution clearly also serves as an 
upper bound of the weights \rhosdn at a given energy. 

The upper panel in \Fig{fig:nrgtruncDd} shows the distribution 
$\nu(E)$ in \Eq{nu:E} of the energies \Esn plotted in the main panel 
(matching horizontal axis). This distribution shows a strong increase 
with energy $E$, consistent with the notion that the many-body phase 
space grows quickly as the available energy for excitations becomes 
larger. Towards large energies, eventually, the data is necessarily 
truncated to the finite number \Nkept of kept states, which leads to 
a drop in the density $\nu(E)$. The exact boundary with respect to 
energy is somewhat blurred, though, since in given case fixed \Nkept 
allows the energy range to vary for different iterations $n$. 
The right panel of \Fig{fig:nrgtruncDd}, on the other hand, shows the 
distribution $\nu(\rho)$ in \Eq{nu:rho} of the weights \rhosdn 
plotted in the main panel (matching vertical axis). This distribution 
is peaked around the largest weights \rhosdn for the largest energies 
\Esn.

The data in the main panel of \Fig{fig:nrgtruncDd} is typically 
bunched around a set of energies for a fixed iteration $n$. This is 
also reflected in the distribution $\nu(E)$ in the upper panel of 
\Fig{fig:nrgtruncDd}, and is due to the discretization of the model. 
Moreover, two iterations are highlighted in strong colors. These 
correspond to the iterations whose energy range is smallest 
($\nmin=6$, red bullets) or largest ($\nmax=74$, black diamonds). 
Intuitively, the largest numerical error is expected from iterations 
such as \nmin (red bullets) since through \Eq{eq:rhoE-decay}, 
stopping at premature energies directly translates to largest 
missing, \ie \emph{discarded} weight in the density matrix. As an 
aside, this serves as a strong argument in favor of truncation \wrt a 
fixed energy cutoff \Ekept rather than a fixed number \Nkept of 
states. Fixed \Ekept, however, also introduces more noise to the data 
in particular for higher lying states. Hence both truncations will be 
used and pointed out in context. 

The weights $\rhosdn$ in the main panel of \Fig{fig:nrgtruncDd} show 
significant vertical spread, which translates into a pronounced tail 
towards exponentially smaller $\rho$ in the distribution $\nu(\rho)$ 
in the right panel. For a given energy $E$ therefore, many of the 
states have \emph{order of magnitudes} lower weight than the top-most 
weights close to $\rho(E)$ in the main panel. This indicates that the 
energy representation with its corresponding diagonal weights \rhosdn 
is not necessarily the optimal basis to analyze accuracy. Moreover, 
note that using the energy eigenbasis \snr with energies \Esn in the 
analysis of the reduced density matrices, this actually mingles the 
energy scales of an effectively larger system $\hat{H}_{n+\dn}$ with 
the basis generated \wrt $\hat{H}_{n}$ only. 

\begin{figure}[tb]
\begin{center}
\includegraphics[width=1\linewidth]{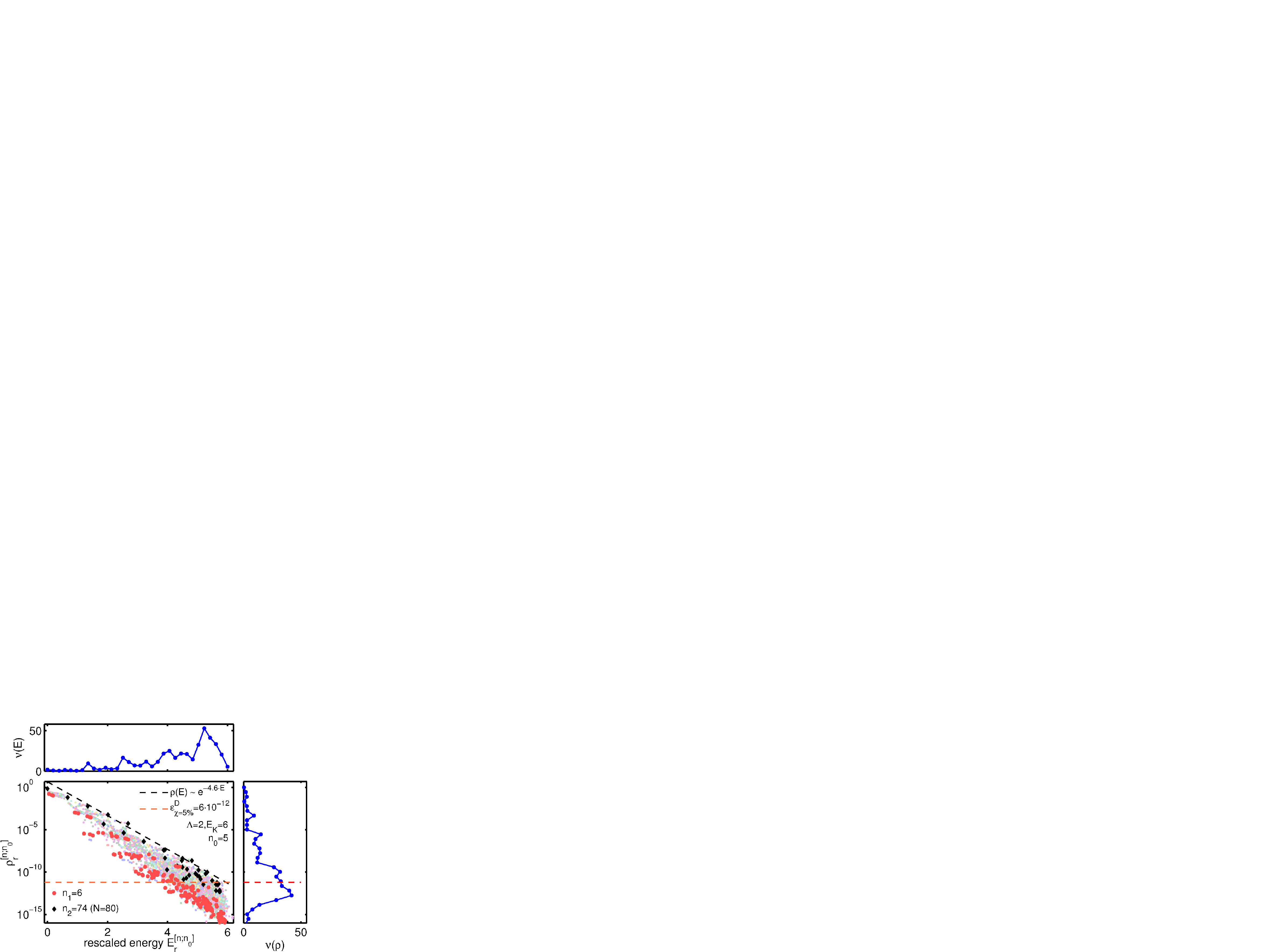}
\end{center}
\caption{
(Color online) Similar analysis as in \Fig{fig:nrgtruncDd} (see 
caption there for further information) for the same underlying 
Hamiltonian, except that the eigenspectrum of the reduced density
matrices in \Eq{eq:rho-eig} was used together with \Eq{eq:reig}
and a fixed energy cutoff $\Ekept=6$. Similar to 
\Fig{fig:nrgtruncDd}, only those iterations are shown where 
truncation occurred (same color for data from the same iteration), 
with the same two iterations highlighted as in \Fig{fig:nrgtruncDd}, 
indicated by $n_1$ and $n_2$. The estimate for the overall discarded 
weight $\epsD_{\chi=5\%} \simeq 6\cdot 10^{-12}$ as defined in \Eq{eq:NRG-epsD} 
is indicated by the horizontal dashed line. 
}%
\label{fig:nrgtruncEe}
\end{figure}

\subsubsection{Eigenbasis of reduced density matrices}

From the point of view of a variationally optimal representation of 
the ground state space of an enlarged system, on the other hand, one 
is directly led to the \emph{eigenspectrum} of the reduced density 
matrix, as exemplified within DMRG. \cite{White92}
The analysis of \Fig{fig:nrgtruncDd} therefore is repeated for the 
same underlying Wilson chain, yet with two modifications: (i) the 
eigendecomposition of the reduced density matrices in \Eq{eq:rho-eig} 
together with \Eq{eq:reig} is used instead of the energy eigenbasis, 
and furthermore (ii) the NRG truncation criterion is based on a fixed 
energy cutoff, $\Ekept=6$. The results are shown in 
\Fig{fig:nrgtruncEe}, with striking quantitative differences compared 
to \Fig{fig:nrgtruncDd}. The spread in the scatter plot is 
significantly narrowed, and overall, the data decays much faster with 
$\kappa \simeq 4.6$, \cf \Eq{eq:rhoE-decay}. Therefore this leads to 
a clearly improved separation of the actually relevant states for the 
subsequent description of the lower-energy scales. This suggests that 
many of the NRG eigenstates, as their energy increases, loose 
importance much faster as compared to \Fig{fig:nrgtruncDd}, despite 
the relatively large diagonal weights $\rho_s$ in the density matrix 
still seen there. In a sense, the weights there represent mere 
matrix-elements in a non-diagonal representation. 

The iterations highlighted in \Fig{fig:nrgtruncEe} are the same 
iterations as in \Fig{fig:nrgtruncDd}. Given a fixed energy cutoff 
$\Ekept=6$ here, however, both have a comparable energy range (hence 
the altered notation $n_1$ and $n_2$), with the number \Nkept of kept 
states varying from $\sim 1000$ at very early iterations (in 
particular iteration $n_1$), down to $\sim 250$ at late iterations 
(such as iteration $n_2$). Note also the markedly fewer data points 
seen for iteration $n_2$. This is only partly due to the reduced 
number of states, as there are also large systematic (approximate) 
degeneracies at the strong-coupling Kondo fixed point already reached 
at this iteration. This results in many of the black diamonds lying 
indistinguishably on top of each other (see also discussion on 
entanglement spectra later). 

As seen from above discussion, rather than taking the energy 
eigenstates \snr and the corresponding diagonal matrix elements 
\rhosdn (\Fig{fig:nrgtruncDd}), the eigenvalues \rhordn of the 
reduced density matrix \rhondn do represent a clearly better choice 
for the analysis of accuracy or entanglement in the system 
(\Fig{fig:nrgtruncEe}), and thus will be used henceforth. This 
prescription also shows a more systematic exponential decay all the 
way down to numerical double precision noise ($10^{-16}$), with the 
decay rate $\kappa$ of $\rho(E)$ roughly independent of the 
discretization parameter $\Lambda$.

\begin{figure}[tb]
\begin{center}
\includegraphics[width=0.95\linewidth]{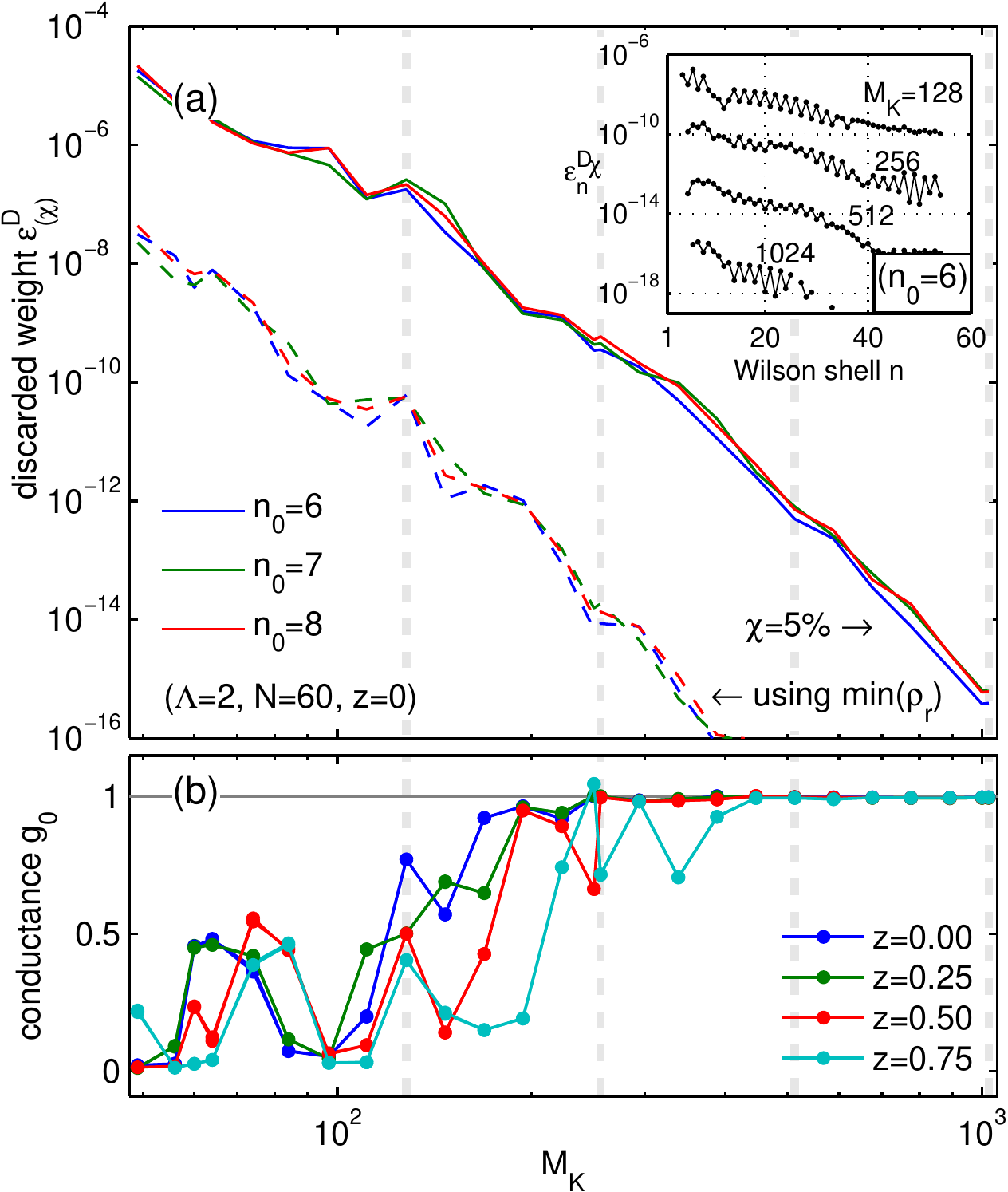}
\end{center}
\caption{
(Color online) Discarded weight \epsD for the SIAM [\Eq{eq:H-SIAM}: 
$U=0.20,\ \epsilon_{d}=-U/2,\ \Gamma =0.01$ (same parameters as in 
\Fig{fig:nrgtruncDd}), with $T_K \simeq 1.23\cdot 10^{-5}$]. 
Panel (a) shows the discarded weight $\epsDX$ defined in 
\Eq{eq:NRG-epsD} vs. \Nkept using $\dn \in\{6,7,8\}$. The data 
$\epsD_{\chi=5\%}$ is shown in solid lines, while the data 
based on the minimum eigenvalue of \rhondn (\cf \Eqt{eq:epsMnn0})
is shown in 
dashed lines. The distribution of the discarded weight $\epsDc_n$ 
along the Wilson chain is shown in the inset for $\Nkept \in 
\{128,256,512,1024\}$, also marked by the vertical dashed lines in 
the main panels. 
Panel (b) shows the conductance $g_{0}$ vs. \Nkept in units of 
$2e^{2}/h$ while using a set of shifted discretizations, with the 
z-values as specified. Convergence in the conductance towards the 
expected unitary limit is seen for $\Nkept\gtrsim 400$, \ie 
$\epsD_\chi\lesssim 10^{-12}$. %%\vspace{-0.2in}
}%
\label{fig:nrgtrunc_N}%
\end{figure}

\subsubsection{Definition of discarded weight}

With the motivation above, the definition of the discarded weight is 
based on the eigendecomposition of the reduced density matrices 
\rhondn in \Eq{eq:rho_nn0}, using the combined data of 
\Eq{eq:rho-eig} and \Eq{eq:reig}. In terms of \Fig{fig:nrgtruncEe}, 
adding more states to the calculation essentially extends the data to 
larger energies and smaller weights, while the large-weight 
low-energy sector already remains widely intact. Therefore the 
largest discarded weight, \ie the weight missing by states not 
included and hence not available, can be estimated, to a good 
approximation, up to an overall prefactor by the smallest weights in 
the kept state space, which are easily accessible. Given the 
exponential decay of the weights together with the residual spread in 
the data as seen in \Fig{fig:nrgtruncEe}, the discarded weight at 
given iteration $n$ can thus be defined through the \emph{average} 
weights \rhordn for the highest energies \Erdn in the kept space, 
\begin{subequations}
\begin{align}
  \epsKn & \equiv
  \Bigl\langle \rhordn \Bigr\rangle _{\Erdn \geq (1-\chi) \max(\Erdn)}
\text{.}\label{eq:epsKnn0}
\end{align}
The parameter $\chi \ll 1$ is considered small, yet is chosen large 
enough (typically $\chi \simeq 0.05$) to average over the residual 
spread of weights. Alternatively and for comparison, an even simpler 
measure in terms of the minimum eigenvalue of \rhondn will be 
considered, 
\begin{align}
  \epsXn & \equiv \min\left( \rhordn \right)
\text{,}\label{eq:epsMnn0}
\end{align}
\label{eq:epsXnn0}
\end{subequations}
which no longer makes any explicit reference to energies. Note that 
even though \epsKn or \epsXn, written \epsKx in short, are purely 
determined within the kept space, they clearly represent a sensible 
estimate for the discarded weight at iteration $n$, \ie $\epsDx_n 
\sim \epsKx$, defined as the fraction of relevant state space missing 
from the latter description of the low energy physics. If no 
truncation has occurred at iteration $n$, however, such as typically 
for the first $n<\dn$ iterations, of course, then there is no 
truncation error either, hence $\epsDx_n=0$ for these iterations. 

In summary, the discarded weight \epsDn at iteration $n$ is defined 
as follows, 
\begin{equation}
   \epsDx_{n} \equiv \left\{
   \begin{array}
     [c]{ll}%
     \epsKx & \text{in the presence of truncation}\\
     0 & \text{without truncation at iteration $n$.}
   \end{array}\right.
\label{eq:NRG-epsD}%
\end{equation}
Here \epsKx can be determined efficiently by including and analyzing 
$\dn$ further NRG iterations within the kept space, where typically 
$\dn \ll N$, cf. \Eq{eq:def_n0}. The overall discarded weight \epsDX 
of a full NRG run then is taken, for simplicity, as the largest 
discarded weight per iteration, 
\begin{equation}
   \epsDX \equiv \max_{n}\left(  \epsDx_{n}\right)
\text{,}\label{eq:NRG-epsD}%
\end{equation}
Using $\chi=5\%$ as in \Eq{eq:epsKnn0}, the discarded weight for the 
NRG run in \Fig{fig:nrgtruncEe} is estimated by $\epsD_\chi \simeq 
6\cdot 10^{-12}$, indicated by the horizontal dashed line. As seen 
from \Fig{fig:nrgtruncEe}, the overall discarded weight $\epsD_\chi$ 
for an NRG run essentially coincides with $\rho(E)$ at the largest 
energies within the kept space. On the other hand, $\epsD$, \ie 
without the usage of $\chi$ based on the plain \emph{minimum} 
eigenvalue of the reduced density matrices \rhondn, \cf 
\Eq{eq:epsMnn0}, will in general lie a (constant) few orders of 
magnitude lower, as it happens, for example, for the data in 
\Fig{fig:nrgtruncEe}. Nevertheless, as will be shown in the 
following, up to an overall global prefactor the discarded weight 
based on either, \epsD or $\epsD_\chi$, both behave in an essentially 
similar fashion. 

\subsection{Application}

The discarded weight \epsDX defined in \Eq{eq:NRG-epsD} sensitively 
depends on the number \Nkept of states kept or the energy threshold 
\Ekept. From \Fig{fig:nrgtruncEe} one expects a strongly diminishing 
discarded weight with increasing \Nkept or \Ekept, a quantitative 
analysis of which is presented in 
\Figs{fig:nrgtrunc_N}{fig:nrgtrunc_E} for the SIAM. 
Figure \ref{fig:nrgtrunc_N} analyzes the dependence of the discarded 
weight \epsDX on the number \Nkept of states kept. As seen in panel 
(a), the discarded weight \epsDX strongly decays with \Nkept, with 
minor variations when a new Wilson shell is fully included without 
truncation, \eg at $\Nkept \in \{256,1024\}$. With panel (a) being a 
log-log plot, the decay of the discarded weight with \Nkept rather 
resembles a polynomial convergence, yet with very large power (on the 
order of 10). The reason for the slower than exponential decay is due 
to the strong increase in the density of states $\nu(E)$ of the full 
many-body eigenspectrum with increasing $E$ as discussed with 
\Figs{fig:nrgtruncDd}{fig:nrgtruncEe}. 

Together with the analysis of the discarded weight in 
\Fig{fig:nrgtrunc_N}, an independent physical check for convergence 
is provided by the numerically computed conductance $g_{0}$ in units 
of $2e^{2}/h$ shown in \Fig{fig:nrgtrunc_N}(b). The conductance was 
calculated via the (spin-resolved) spectral function 
$A_{(\sigma)}(\omega)=\int \tfrac{dt}{2\pi} e^{i\omega t}\langle 
\{\hat{d}_\sigma^{\phantom\dagger}(t), \hat{d}_\sigma^\dagger 
\}\rangle_T$ of the impurity level, with $g_{0} = \pi\Gamma\int 
d\omega (-\tfrac{\partial f}{\partial\omega}) A(\omega)$. Here the 
Fermi function $f(\omega)$ and the spectral function $A(\omega)$ are 
evaluated at small but finite temperature $T\simeq 6\cdot10^{-8}$, 
which is much smaller than the Kondo temperature of $\TK \simeq 
1.23\cdot 10^{-5}$ for given parameter set and corresponds to the 
energy scale close to the end of the Wilson chain, having $\Lambda=2$ 
and $N=60$. 
Expecting $g_{0}=1$ for the symmetric SIAM, the data in 
Fig.~\ref{fig:nrgtrunc_N} indicates convergence for $\Nkept \gtrsim 
400$. The data for smaller \Nkept is not yet converged, and therefore 
(strongly) depends on numerical details, such as non-averaged 
z-shifts. \cite{Oliveira92,Zitko09} 

With \Nkept being constant, the energy of the topmost kept states can 
vary significantly with Wilson shell $n$, which directly also leads 
to a clear dependence of the discarded weight \epsDX on $n$. This is 
shown in the inset to panel (a) for the set of different values of 
\Nkept marked in the main panels by the vertical dashed lines. The 
discarded weight $\epsD_\chi$ clearly varies over more than three 
orders of magnitude within a single NRG run, irrespective of the 
actual \Nkept. In particular, one can see that earlier iterations 
dominate the discarded weight $\epsD_\chi$ for physical reasons. In 
the strong-coupling regime for $n \gtrsim \nK$ (with iteration $\nK 
\simeq 35$ corresponding to the energy scale of $\TK$), the discarded 
weight is smallest, while for the intermediate free orbital or local 
moment regime for $n \lesssim \nK$, these regimes require a 
\emph{larger} number of states for comparable numerical accuracy from 
a physical point of view, indeed. 

\begin{figure}[tb]
\begin{center}
\includegraphics[width=0.96\linewidth]{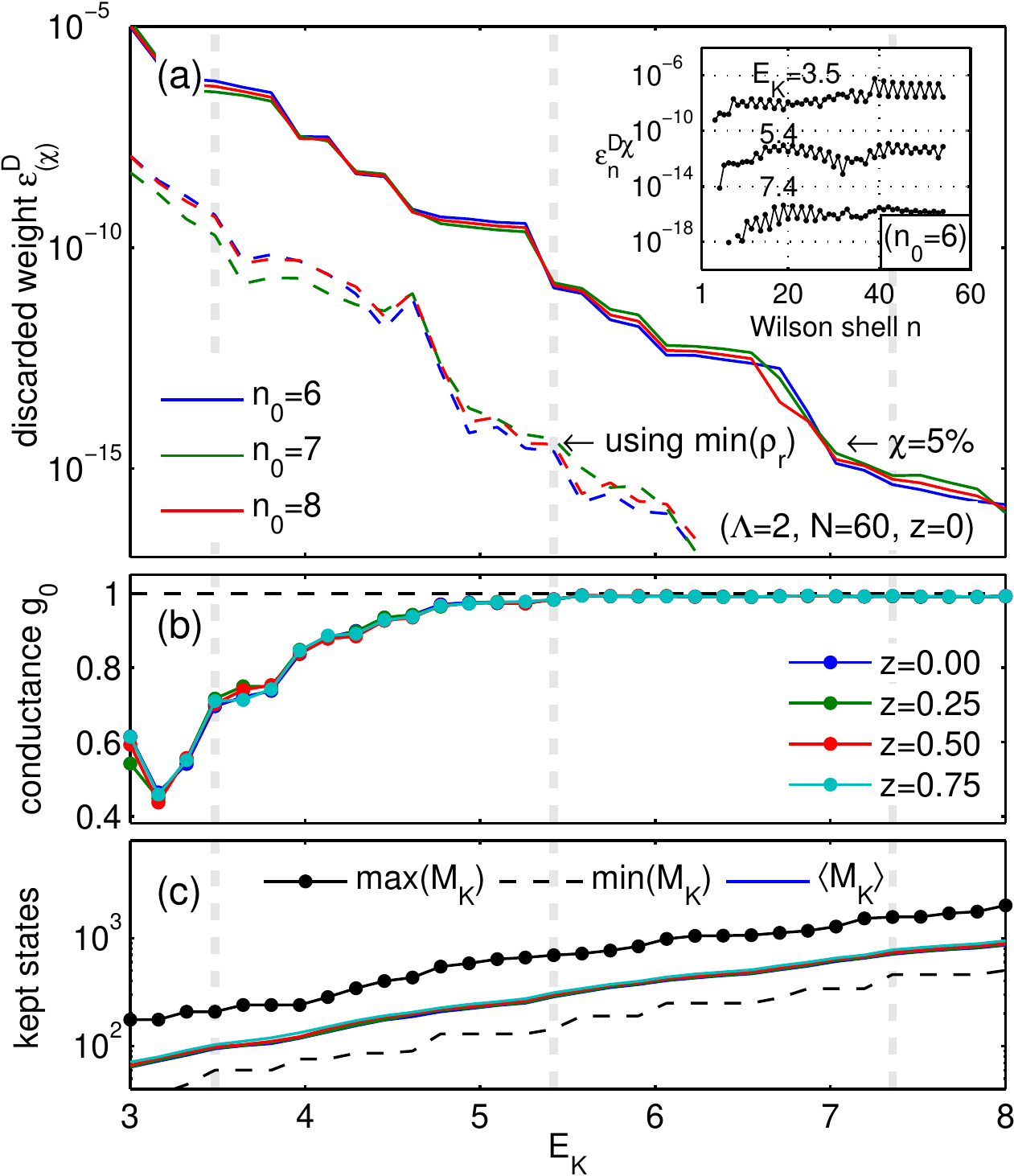}
\end{center}
\caption{
(Color online) Similar analysis as in \Fig{fig:nrgtrunc_N},
yet for truncation with respect to fixed energy \Ekept.
For several values of \Ekept marked by the vertical dashed lines
in the main panels, the distribution of the discarded weight
$\epsDc_n$ along the Wilson shell $n$ is shown in the inset to panel (a).
With \Nkept allowed to vary over a wider range, panel (c) shows 
the correlation of \Nkept with \Ekept, plotting average, minimum, and 
maximum of \Nkept along the Wilson chain. For the average \Nkept,
data for different z-shifts is shown (several lines
on top of each other, with same color coding as in panel b).
\vspace{-0.2in}
}%
\label{fig:nrgtrunc_E}%
\end{figure}

Given the underlying energy scale separation of the NRG, a 
straightforward way to obtain a more equally distributed $\epsDx_n$ 
is achieved using an energy cutoff \Ekept, as demonstrated in 
\Fig{fig:nrgtrunc_E} for exactly the same system as in 
\Fig{fig:nrgtrunc_N} otherwise. For the values of \Ekept indicated by 
the vertical dashed lines in the main panels, the inset to 
\Fig{fig:nrgtrunc_E}(a) shows the distribution of $\epsDc_n$. By 
construction, the discarded weight is, up to even-odd oscillations, 
clearly more uniformly distributed over the Wilson shells as compared 
to the case of fixed \Nkept in \Fig{fig:nrgtrunc_N}(a). The discarded 
weight in panel (a) clearly diminishes exponentially with \Ekept, yet 
with pronounced intermediate plateaus since the discrete 
eigenenergies within an NRG run are usually bunched around certain 
energies. The corresponding average \Nkept as function of \Ekept, 
nevertheless, follows a rather smooth monotonic behavior, as shown in 
panel (c). Given fixed \Ekept, however, clear variations of \Nkept 
are seen within a given NRG run, hence also smallest and largest 
\Nkept are shown in panel (c). Ignoring iterations without 
truncation, in given example, typically the largest \Nkept is 
required at early iterations, while the smallest \Nkept are 
encountered in the strong coupling regime at late iterations $n 
\gtrsim n_K$. 

The calculated conductance shown in panel (b) converges clearly more 
uniformly with increasing \Ekept as compared to 
\Fig{fig:nrgtrunc_N}(b). In particular, it indicates converged NRG 
data for $\Ekept \gtrsim 5.5$, which corresponds to $\epsD_\chi 
\lesssim 10^{-12}$. Therefore in both settings, for constant \Nkept 
in Fig.~\ref{fig:nrgtrunc_N} as well as for constant \Ekept in 
Fig.~\ref{fig:nrgtrunc_E}, convergence of the physical data is found 
for a similar discarded weight of $\epsD_\chi \lesssim 10^{-12}$ with 
a negligible dependence on \dn. This value therefore is considered a 
sufficient bound in accuracy to capture the main physics, with other 
quantities such as the NRG energy flow diagram already also well 
converged. 

Alternatively, using the plain minimum of the eigenvalues of the 
reduced density matrices in \Eq{eq:epsMnn0}, this leads to 
convergence for $\epsD \lesssim 10^{-16}$. Given that $\epsD$ refers 
to the minimum eigenvalue in the kept space, $\epsD$ consistently 
lies about three orders of magnitudes lower than $\epsD_\chi$ and is 
considered a lower bound to the actual discarded weight. While \epsD 
fluctuates slightly more strongly compared to $\epsD_\chi$ owing to 
the fact that it is not an averaged quantity such as $\epsD_\chi$, it 
nevertheless follows a similar consistent picture in terms of 
convergence with the number \Nkept of states kept or the energy 
\Ekept used for truncation. 
In this sense, either discarded weight, $\epsD$ as well as 
$\epsD_\chi$, can be used quite generally as a quantitative measure, 
indeed, to demonstrate accuracy within the NRG. In order to avoid 
confusion, however, it shall be made clear which one is used. 

\section{Entanglement spectra}

% The averaged lower end of the spectral decomposition of the reduced 
% density matrices \rhondn in \Eq{eq:rho_nn0} was used to estimate the 
% discarded weight in \Eq{eq:epsKnn0} above. There, some fixed small 
% \dn was sufficient to prove that within that short range, the 
% truncation does not compromise the description of the subsequent low 
% energy sector. Moreover, it was seen from numerics that the discarded 
% weight is rather insensitive to further increase of \dn. 
%
The reduced density matrices \rhondn clearly also carry 
\emph{physical} information in terms of entanglement along the Wilson 
chain. This is provided by the high end of their spectral 
decomposition. There the exact details of the largest eigenvalues of 
\rhondn are of interest, which do vary with \dn over a wider range 
depending on the underlying physics. Hence, in the following, the 
actual entanglement spectra will be calculated with respect to the 
reduced density matrices \rhon of the overall ground state of the 
system, 
\begin{equation}
    \rhon \equiv \lim_{\dn\to\infty} \rhondn
    \simeq \hat{\rho}^{[n;N-n]}_{0}
\text{.}\label{eq:rhon0}
\end{equation}
The length $N$ of the Wilson chain is taken sufficiently large, such 
that the energy scale of the last iteration $N$ is much smaller than 
any other energy scale in the system. Temperature is therefore 
essentially zero. For comparison, also the \emph{truncated} 
entanglement spectra will be calculated from \rhondn for finite small 
\dn, with \dn specified in context. Motivated by the discussion 
following \Eq{eq:FDM-rho}, the latter analysis can be linked to 
finite temperature settings. 

\begin{figure}[tb!]
\begin{center}
\includegraphics[width=\linewidth]{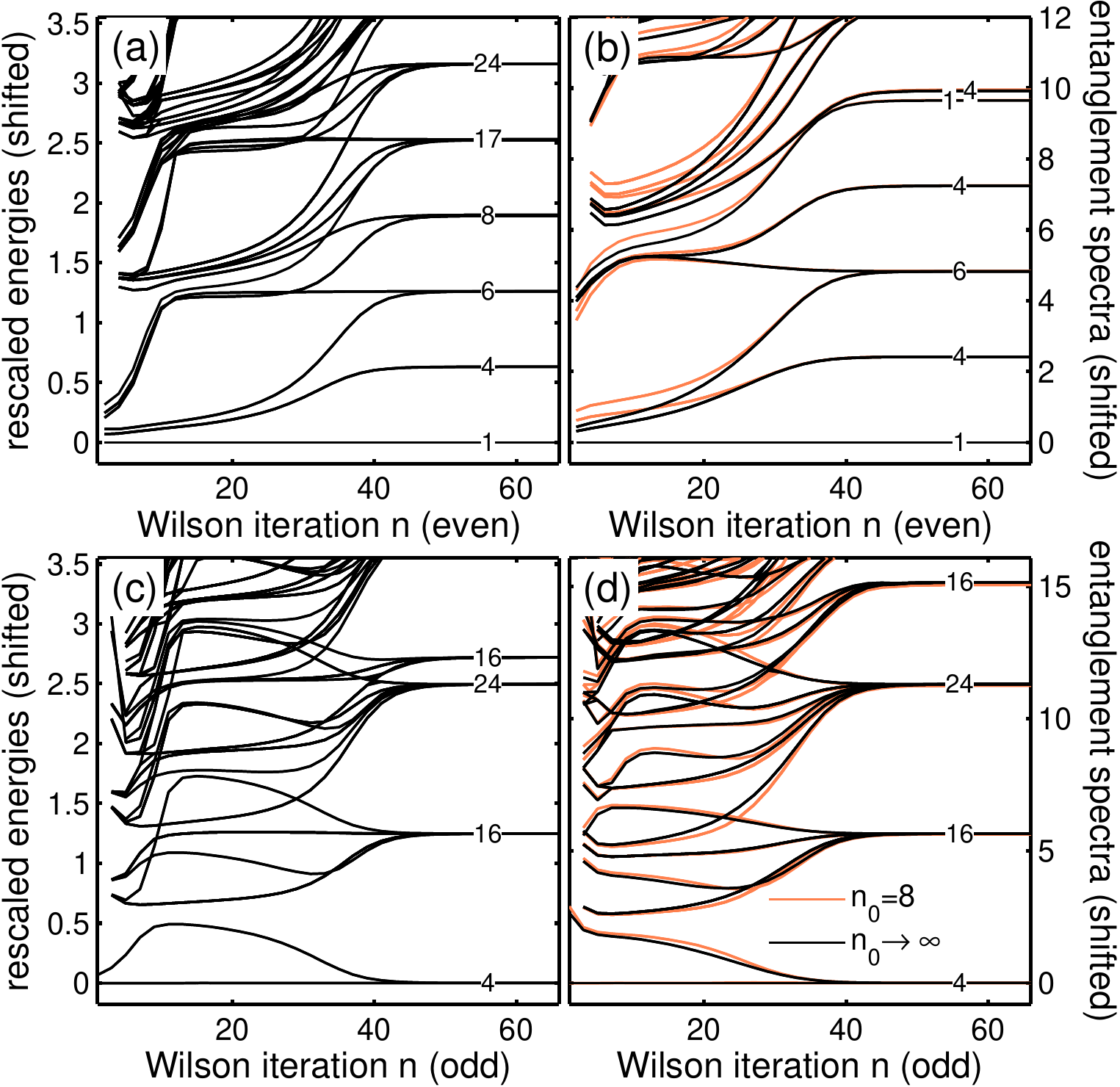}
\end{center}
\caption{ (Color online) 
Comparison of the standard NRG energy flow diagram (left panels)
to the \emph{entanglement flow diagram} (right panels) for the 
symmetric SIAM [$U=0.2$, $\varepsilon_{d}=-U/2$, $\Gamma=0.01$, 
$\TK=1.2\cdot10^{-5}$; $\Lambda=2$, $\Nkept=512$, $N=80$], with top 
(bottom) panels for even (odd) iterations, respectively.
In addition to the actual entanglement flow diagram obtained from 
the ground state of the last iteration at $N=80$ (black lines),
also the truncated entanglement flow diagram is shown, using $\dn=8$ 
(orange (gray) lines). 
For better comparison with the energy flow diagram, the entanglement 
spectra (right panels) are also shifted at every iteration with 
respect to the smallest entanglement energy $\min(\xi)$. The y-scale 
of the entanglement spectra was adjusted to best match the energy 
fixed point spectrum in the left panels. Degeneracies of energies at 
large $n$, \ie lines lying indistinguishably on top of each other, 
are specified by the numbers on top of the lines in all panels. 
}%
\label{fig:entangle-spec}%
% \end{figure}
%
% \begin{figure}[tb!]
\begin{center}
\includegraphics[width=\linewidth]{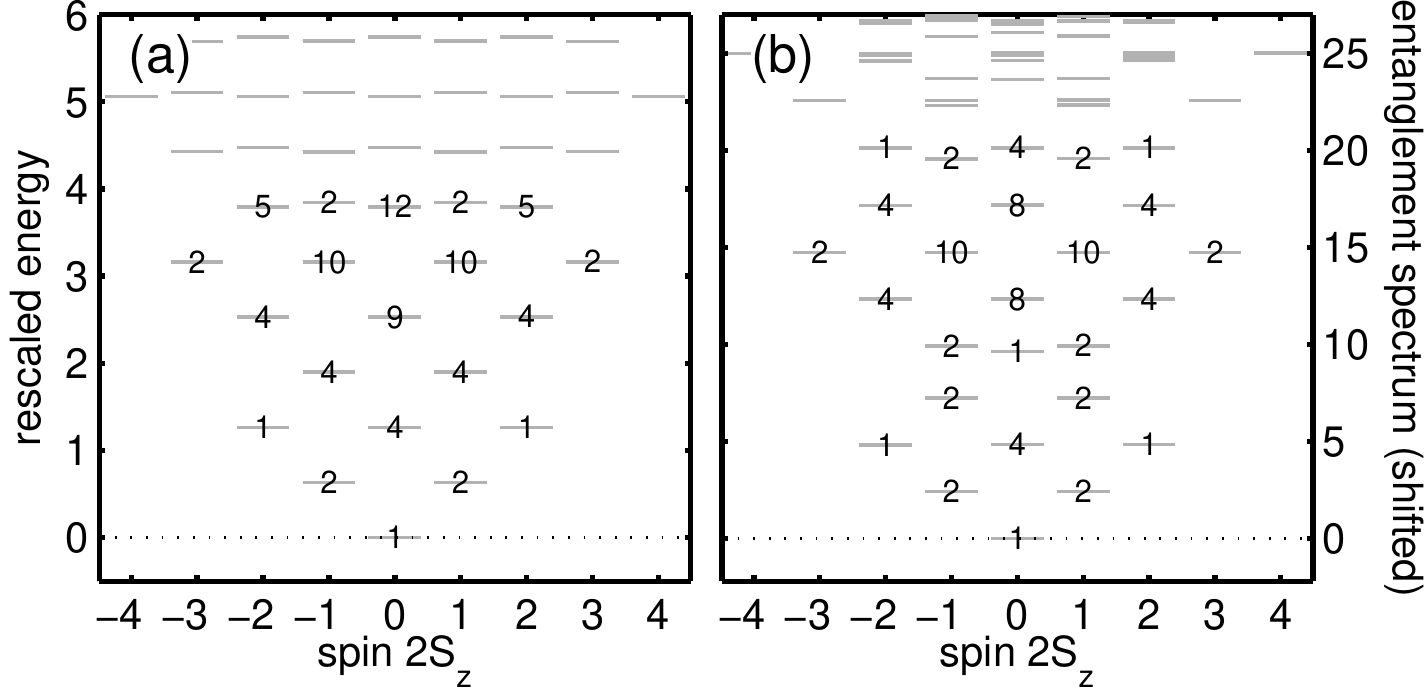}
\end{center}
\caption{ (Color online) Comparison of spin-resolved fixed point 
spectra for the symmetric SIAM in \Fig{fig:entangle-spec} in the SC 
regime ($n=60$). Panel (a)[b] show the energy [entanglement] fixed 
point spectrum, respectively, \vs spin symmetry quantum number $S_z$. 
For all low-energy multiplets the underlying (approximate) degeneracy 
is indicated. The entanglement spectrum is shifted \wrt to its lowest 
energy and scaled to match the energy fixed point spectrum in panel 
(a). \vspace{-0.6in} 
} \label{fig:ESpec:SCs}%
\end{figure}

\subsubsection{General definition}

The partitioning of the Wilson chain into two parts, the chain up to 
and including site $n$ (part A), and the traced out remainder of the 
system (part B) is generic. In particular, this allows to make use of 
the recently introduced entanglement spectra (ES) \cite{Li08} for the 
physical characterization of a given wave function. Here these 
entanglement spectra provide a powerful tool for the systematic 
analysis of the physical correlations in the reduced density matrices 
\rhon in \Eq{eq:rhon0}. 

Consider a given wave function of a some system partitioned into 
parts A and B. The reduced density matrix $\hat{\rho}_{A} \equiv 
\mathrm{tr}_B(\rho)$ is obtained by tracing out part $B$ of the 
overall density matrix $\rho$. Within this setting, the entanglement 
spectrum is defined as the spectrum of the fictitious Hamiltonian 
$\hat{H}_{\rho}^{A}$, \cite{Li08} 
\[
   \hat{\rho}_{A}=:\exp(-\hat{H}_{\rho}^{A}) \text{.}
\]
One may assume an effective inverse temperature $\beta:=1$ in order 
to make contact with a thermal density matrix. This $\beta$ also sets 
the (otherwise arbitrary) energy scale in the \emph{per se} 
dimensionless $\hat{H}_{\rho}^{A}$. With $\hat{\rho}_{A}$ a positive 
operator, the entanglement spectrum $\xi_{r}$ is defined as the 
eigenvalues of $\hat{H}_{\rho}^{A}$, \ie 
\begin{equation}
   \xi_{r}:=-\log\rho_{r}
\text{,} \label{eq:ES-def-logrho}%
\end{equation}
with $\rho_{r}$ the spectral decomposition of the reduced density 
matrix $\hat{\rho}_{A}$. Particular information can be read off from 
the entanglement spectrum as soon as there is a rich amount of 
quantum numbers specifying the entanglement levels and when 
entanglement gaps appear which separate a low-lying generic set of 
levels from irrelevant background correlations. 
\cite{Li08,Thomale10,Thomale10b} The spectra $\rho_{r}$ and $\xi_{r}$ 
are independent of whether A or B is traced out, while of course, 
they are dependent on the specific choice of the partitioning. 
For entanglement spectra, the partitioning typically occurs in real 
space for gapped systems, analyzing the \emph{edge} of the thus 
created boundary, while for gapless systems momentum space is 
preferred. \cite{Thomale10} The second case then is consistent with 
the systematic NRG prescription of energy scales based on the 
underlying discretization in energy (momentum) space. 

By construction, the dominant correlations between systems $A$ and 
$B$ correspond to the lowest \textit{entanglement energies} 
$\xi_{r}$, while weaker correlations will rise to higher energies. By 
tracing out a major part of the system, entanglement spectra provide 
significantly more information, say, than just the entanglement 
entropy between $A$ and $B$. In particular, it has been shown that it 
provides finger prints of the underlying physics, and as such allows 
to characterize the physical nature of a given wave function. 
\cite{Li08,Thomale10} This analysis is therefore entirely targeted at 
a given (ground state) wave function, without any further reference 
to an underlying physical Hamiltonian that it may have originated 
from. 

\subsubsection{Application to NRG}

The general concept of the entanglement spectra can be readily 
transferred to the NRG. At each iteration $n$, the reduced density 
matrix \rhon in \Eq{eq:rhon0} is computed and diagonalized, with its 
eigenspectrum mapped onto the entanglement spectrum in 
\Eq{eq:ES-def-logrho}. Collecting these spectra and plotting them \vs 
iteration index $n$ for even and odd iterations separately, the 
result will be referred to as \emph{entanglement flow diagram}, in 
complete analogy to the standard energy flow diagrams of the NRG. For 
comparison, also the truncated entanglement spectra for finite small 
\dn will be analyzed, which in their combination will be referred to 
as \emph{truncated} entanglement flow diagram. In either case, the 
entanglement spectra are obtained in a \emph{backward} sweep, purely 
based on the iterative low-energy Hilbert-space decomposition of a 
prior NRG run in terms of the A-tensors in \Eq{eq:Atensor}. This is 
in contrast to the energy flow diagram, which is calculated with 
increasing shell index $n$ in a \emph{forward} sweep making explicit 
reference to the Hamiltonian. 

\begin{figure}[tb!]
\begin{center}
\includegraphics[width=\linewidth]{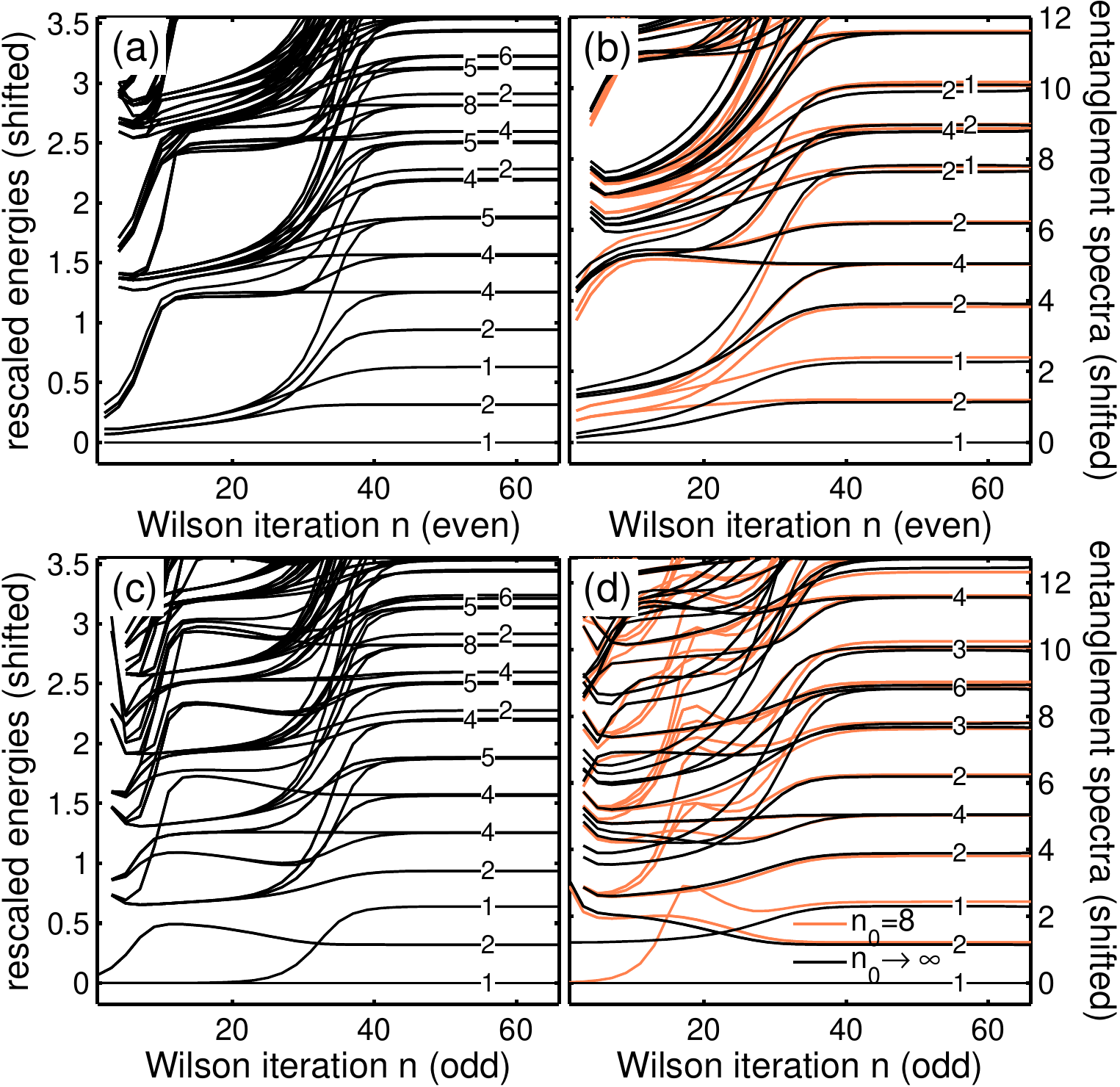}
\end{center}
\caption{ (Color online) 
Comparison of the standard NRG energy flow diagram (left panels) to 
the \emph{entanglement flow diagram} (right panels) for the SIAM at 
finite magnetic field (same analysis as in \Fig{fig:entangle-spec}, 
otherwise, see caption there for details, with same model parameters,
except $B=2\cdot10^{-5} \simeq 1.6\,\TK$). 
}%
\label{fig:entangle-specB}%
%% \end{figure}
%%
%% \begin{figure}[tb!]
\begin{center}
\includegraphics[width=\linewidth]{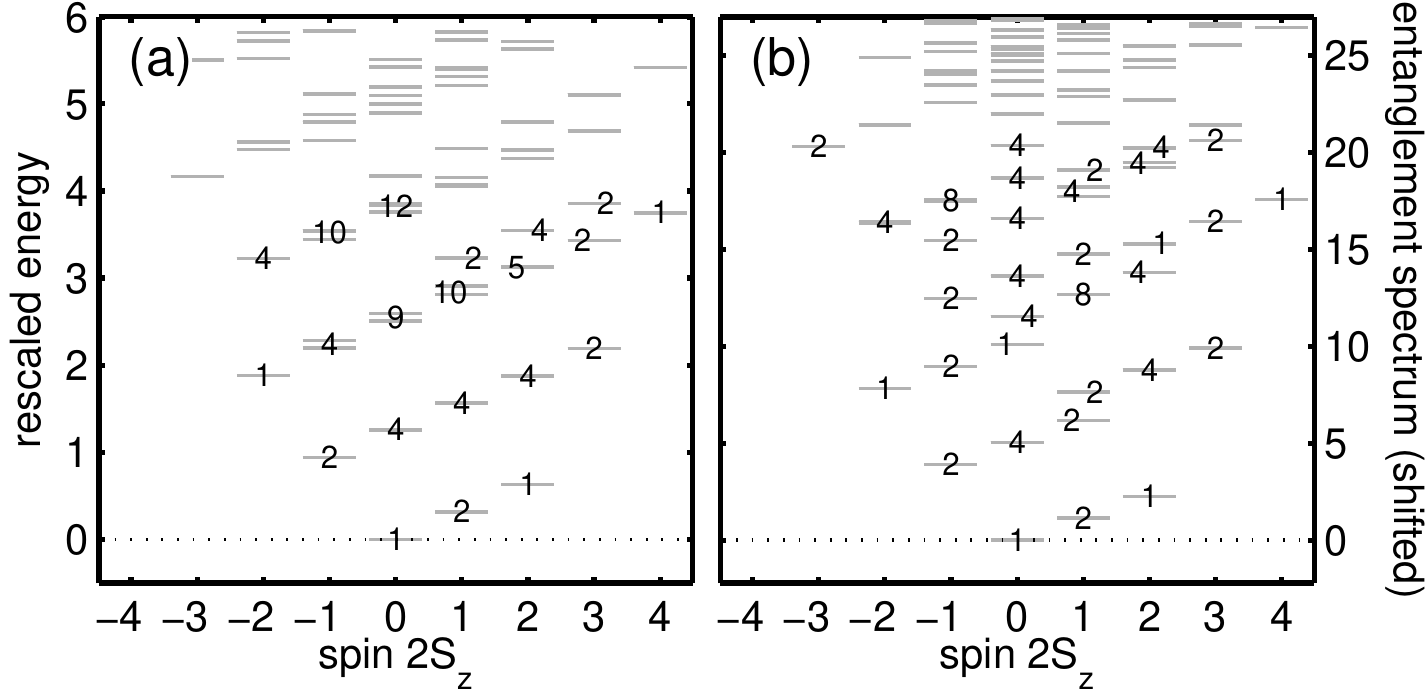}
\end{center}
\caption{
(Color online) Comparison of spin-resolved fixed point spectra for 
the SIAM at finite magnetic field in \Fig{fig:entangle-specB} at the 
even iteration $n=60$ (similar analysis as in \Fig{fig:ESpec:SCs} 
otherwise). 
} \label{fig:ESpec:SCa}%
\end{figure}

The entanglement spectra were calculated for the symmetric SIAM in 
the absence of magnetic field. The resulting entanglement flow 
diagram is presented in \Fig{fig:entangle-spec} together with a 
direct comparison to the standard NRG energy flow diagram. The data 
is plotted for even (odd) Wilson shells $n$ in the upper (lower) 
panels, respectively. The energy flow diagram, shown in the left 
panels, clearly distinguish the well-known physical regimes of the 
SIAM, namely the free orbital regime (FO; $n\lesssim10$), the local 
moment regime (LM; $10\lesssim n\lesssim\nK$), and the strong 
coupling regime (SC; $n\gtrsim \nK$), where $\nK\simeq35$ corresponds 
to the energy scale of the Kondo temperature $\TK=1.2\cdot10^{-5}$, 
having $\Lambda=2$. All degeneracies for $n>\nK$ are explicitly 
specified in \Fig{fig:entangle-spec}. In particular, for even 
iterations, the ground state is unique throughout, \eg the Kondo 
singlet for $n>\nK$ (panel a), while for odd-iterations the ground 
state space at small energies is four-fold degenerate due to the 
particle-hole symmetric parameter set (panel c). 

Interestingly, a very similar picture emerges from the entanglement 
flow diagram in right panels, \Fig{fig:entangle-spec}(b+d) (black 
lines). For comparison, also the truncated entanglement spectra are 
shown using $\dn=8$ (orange (gray) lines), which in given case 
converge rapidly, in fact exponentially, with increasing 
$\dn\lesssim10$ towards the actual entanglement flow diagram. The 
latter then mimic the energy flow diagram in the left panels over a 
wide range. For example, the convergence towards the Kondo fixed 
point occurs around similar iterations, and even the degeneracies of 
the lowest states of the energy flow diagram are exactly recovered by 
the entanglement spectra. The latter also holds on the 
symmetry-resolved level, as demonstrated in \Fig{fig:ESpec:SCs} for 
the even iteration $n=60$ (see later discussion). 
Nevertheless, looking more closely, a few notable qualitative 
differences of the entanglement flow diagrams in right panels of 
\Fig{fig:entangle-spec} are seen compared to the energy flow diagrams 
in the left panels. Overall, the entanglement flow diagrams appear 
shifted by about 5 iterations to larger energies. This can be 
understood, considering that the entanglement spectra are calculated 
for enlarged systems together with the rapid convergence with 
increasing \dn in given case. However, there are further pronounced 
differences with the energy flow diagram for the earliest iterations 
in the FO regime, $n\lesssim10$. 

% -------------------------------------------------------------------- %

These latter differences in the entanglement flow diagram can be 
significantly enhanced by turning on a magnetic field on the order of 
the Kondo temperature, as shown in \Fig{fig:entangle-specB} for 
$B=1.6\,\TK$. This corresponds to the energy scale at iteration 
$\nB\simeq 32$, given $\Lambda=2$. The magnetic field has been chosen 
such, that for late iterations $n\gg \nB$ the fixed point spectrum 
for even and odd iterations become essentially the same (compare the 
low-energy fixed point spectra in panels a(b) to c(d), respectively). 
Due to the magnetic field, the Kondo singlet (previously the unique 
state at even iterations) is largely destroyed for $n\gtrsim\nB$ with 
a sizeable magnetization at the impurity. Clearly, the NRG eigenbasis 
at early iterations $n<\nB$ does not yet know about the small energy 
physics to come (\eg the small $B\sim \TK$ applied in given case). 
Therefore the energy flow diagram essentially remains unaltered 
there, when compared to the case without magnetic field in 
\Fig{fig:entangle-spec}(a+c). The flow changes strongly only starting 
from the energy scale of the magnetic field value, \ie for $n>\nB$ 
where it moves into a different fixed-point spectrum. In particular, 
there also emerges a unique state now in the energy flow diagram for 
odd iterations for $n\gg\nB$, \ie the symmetry broken spinful state 
favored by the magnetic field. 
By including magnetic field, the entanglement flow diagram shows 
pronounced differences from the energy flow diagram for 
$n\lesssim\nB$, which includes large portions of the LM regime. While 
the energy spectrum up to and including site $n$ is ignorant of the 
low-energy physics to come, this very low-energy physics is captured 
by the reduced density matrices and thus reflected in the 
entanglement flow diagram. \footnote{Note that it was exactly this 
kind of reasoning, for example, that led to the success of 
density-matrix-based NRG methods for dynamical properties, starting 
with DM-NRG. \cite{Hofstetter00} For the SIAM, for example, the 
reduced density matrices, while not crucially important in the 
absence of magnetic field, are absolutely essential for the correct 
description of spectral correlations at finite magnetic field. There 
it is intuitively clear that the spin-resolved spectral function 
$A_\sigma(\omega)$ at the impurity at small temperatures $T\ll\TK$ in 
the presence of a magnetic field $B>\TK$ redistributes significant 
spectral weight at \emph{large} energies $|\omega|\gg\TK$, which 
accounts for the breakup of the Kondo singlet. Within the NRG, this 
thus translates into a \emph{feedback} from small to large energies, 
which is captured correctly only after including the reduced density 
matrices for the remainder of the system. \cite{Hofstetter00} Yet 
contributions from all NRG shells are required to cover the full 
spectral range of dynamical correlation functions. \cite{Costi97} A 
clean combined prescription for this was finally provided by the 
FDM-NRG approach \cite{Wb07} based on complete basis sets. 
\cite{Anders05,Peters06} }

Consider the entanglement spectra derived from the overall ground 
state (black lines) in \Fig{fig:entangle-specB}(b+d). In panel (d) 
the ground state remains unique throughout, \ie remembers the 
symmetry broken magnetic state, determined at much lower energy 
scales, all the way up to the largest energies. Within the split-up 
lowest energy space with subsequent degeneracies [1-2-1] in panel (d) 
for $n\gg\nB$ (to be called [1-2-1] configuration), the first and 
second excited states cross each other with decreasing $n$ leading to 
a [1-1-2] configuration for small $n$, \ie large energies. 
Nevertheless, the singly degenerate excited state clearly remains 
split-off, and does not merge with the ground state, which is in 
strong contrast to the energy flow diagram in panel (c) with a [2-2] 
configuration for $n\ll\nB$. This degeneracy in the ground state 
space that is ignorant of the small magnetic field is partly 
reflected only in the \emph{truncated} entanglement flow diagram. 
Using small \dn (orange (gray) lines in panels d), this eventually 
also misses the low energy physics. Therefore these spectra in panel 
(d) eventually are also in a [2-2] configuration for the smallest 
$n$, with a more irregular transient behavior with increasing $n$. A 
similar trend is also observed for even iterations in panels (a+b). 
While the ground state remains unique for all iterations in both 
panels, the entanglement flow in panel (b) tends to split off the 
excited levels right above the lowest [1-2-1] state space 
configuration for small $n$. For the truncated entanglement flow, on 
the other hand, the lines of these excited levels remain entangled 
with higher excitations, which is similar to the situation in the 
energy flow diagram in panel (a). 

Nevertheless, the low-energy fixed-point spectra for $n\gg \nB$ again 
agree well for both the energy and entanglement flow diagram in 
\Fig{fig:entangle-specB}, which again also holds for the 
symmetry-resolved spectra, as demonstrated for the even iteration 
$n=60$ in \Fig{fig:ESpec:SCa}. This agreement in the spectra of the 
stable low-energy fixed point, present in both the non-magnetic as 
well as the magnetic case, is understood as a generic feature. There 
both, the energy eigenstates as well as the reduced density matrices 
are deeply rooted in the low-energy physics, \ie of the overall 
ground state of the system at $T\to0$, and hence present a consistent 
description of the system. 

The detailed structure of the energy fixed point spectra provides 
clear physical information. \cite{Wilson75,Bulla08} This includes, 
for example, phase shifts if a Fermi-liquid point of view is 
supported as is the case for the SIAM. This then directly explains 
all of the splittings and degeneracies in the low energy sector of 
the energy fixed point spectra. For example, consider the energy 
spectrum in \Fig{fig:ESpec:SCs}(a) for the fully symmetric SIAM in 
the non-magnetic case. Note that while spin-resolved spectra are 
shown in \Fig{fig:ESpec:SCs}, in given case the charge-resolved 
spectra would look exactly the same due to particle-hole symmetry. 
With the spectra shown for an even iteration, the ground state is 
unique, \ie represents the Kondo singlet with $S_z=0$. The first 
excited states for $S_z=+\tfrac{1}{2}$, correspond to an extra 
particle with spin-up or a hole with spin-down. Given particle-hole 
symmetry, both processes have the same energy $\delta/2=0.63$ (in 
rescaled energy units), and hence are two-fold degenerate, indicated 
by the number on top of the level in \Fig{fig:ESpec:SCs}. By 
symmetry, the same excitations exist for $2S_z=-1$, leading to the 
[2-2] degeneracy (4 states) in the lowest excitations in 
\Fig{fig:ESpec:SCs}(a). The next higher excitation combines two of 
above processes. This leads to a total of 6 excitations, all with 
energy $\delta$ and distributed over $2S_z\in\{-2,0,+2\}$. Here two 
of the excitations at $2S_z=0$ correspond to the extraction or 
annihilation of two particles with opposite spin. This fully explains 
the [1-4-1] degeneracy of the excited states at energy $\delta=1.26$ 
in \Fig{fig:ESpec:SCs}(a), and also the combined 6-fold degeneracy 
seen in the energy flow diagram seen at this energy in 
\Fig{fig:entangle-spec}(a). The argument can be continued along 
similar lines to explain the [4-4] (8 states) and [4-9-4] (17 states) 
degenerate subspaces of the next higher excitations. Excitations with 
even higher energy eventually have missing levels due to NRG 
truncation. 

The same analysis as for the energy spectra, however, cannot be 
applied with equal rigor to the entanglement spectra. While the 
ground state [1] and the lowest [2-2] and [1-4-1] excitations in 
\Fig{fig:ESpec:SCs}(b) fully agree in symmetries, degeneracy and also 
in the precise relative level spacing, the next higher [4-4] 
excitation in panel (a) is broken up in \Fig{fig:ESpec:SCs}(b), with 
some of the levels shifting to higher entanglement energy. 
Nevertheless, the degenerate set [2-10-10-2] further up in energy 
still again equally appears for both, energy and entanglement 
spectra. 

The same analysis as in \Fig{fig:ESpec:SCs}, is repeated for the 
magnetic case in \Fig{fig:ESpec:SCa} for the same even iteration 
$n=60$. Despite the rather different level spectrum for large $n$ in 
the flow diagram in \Fig{fig:entangle-specB}, the actual 
spin-resolved fixed point spectrum is qualitatively very similar to 
the non-magnetic case in \Fig{fig:ESpec:SCs}. Aside an overall tilt 
of the level structure, all degeneracies and level positions of the 
lower part of the energy spectrum in panel (a) are again fully 
described by elementary single-particle excitations. The underlying 
reason for this similarity of the fixed points spectra in the 
magnetic and non-magnetic case is that, apart from the (screened) 
impurity spin, the system is well described by an effective 
Fermi-liquid picture. With the low energy fixed point spectra well 
reflected in the entanglement spectra, a similar tilt in the level 
structure is also observed in \Fig{fig:ESpec:SCa}(b) when compared to 
\Fig{fig:ESpec:SCs}(b). Note, for example, that to the lower left of 
the spectrum the same [1-2-1], as well as the [2-4-2, 2-4-2] state 
sequence with increasing energy is seen. 

\section{Summary and outlook}

The reduced density matrices of the NRG by tracing out the low-energy 
sector have been analyzed in detail. The low end of their 
eigenspectra was used to estimate the discarded weight $\epsDX$ in 
\Eqr{eq:epsXnn0}{eq:NRG-epsD} as a quantitative and site-resolved 
measure of the accuracy within the NRG. While, in principle, the same 
reduced density matrices could also be utilized as the basis for an 
altered truncation criteria similar to the DMRG, this, however, 
requires sufficiently large \Nkept to start with. In practice, this 
is sufficiently close to a truncation with respect to an energy 
cutoff \Ekept. Either way, all of this can be easily and quickly 
checked using the proposed analysis in terms of the discarded weight 
which provides a useful quantitative tool. 

Furthermore, the dominant correlations of the reduced density 
matrices were analyzed in terms of their entanglement spectra. Due to 
the NRG flow towards small energy scales, these spectra can be 
combined into entanglement flow diagrams. There different physical 
regimes can be identified similar to the standard NRG energy flow 
diagrams. Considering that the entanglement spectra are obtained 
solely based on the wave function, the agreement of the low-energy 
fixed point spectra are stunning. A possible larger disagreement at 
higher energies, \ie for earlier Wilson shells, on the other hand, 
depends on the specific physical situation. Given the NRG background, 
as an outlook this appears to suggest the following. For all energy 
shells (iterations) $n$ where the entanglement spectrum is 
quantitatively comparable to the NRG energy spectrum for the lowest 
set of states, the reduced density matrices themselves are not 
crucially important in the description of the system. Instead, they 
may be replaced by thermal density matrices in the NRG eigenbasis. In 
a sense, by tracing out the low-energy sector, the resulting reduced 
density matrices maintain an approximate thermal character, with 
implications to thermalization at a given energy shell. 
\cite{Poilblanc11} For energy shells with a qualitative difference 
between the energy and entanglement spectra, however, the reduced 
density matrices are crucially important to capture the correct 
physics in the NRG calculation that explicitly uses data from such 
energy shells. 

A detailed analysis of the deeper connection and the explicit 
differences between the energy and the entanglement spectra appears 
interesting, yet is out of the scope of this paper. In particular, it 
also appears instructive to analyze the entanglement spectra for 
non-fermi liquid systems such as the symmetric two-channel Kondo 
model, as the analysis presented in this paper suggests a strong 
physical connection of the entanglement spectra to the underlying 
physics. 

\begin{acknowledgements}

I want to thank Jan von Delft for a critical review of the script, 
and also Ronny Thomale for helpful comments on entanglement spectra. 
This work has received support from the German science foundation 
(DFG: TR-12, SFB631, NIM, and WE4819/1-1). 

\end{acknowledgements}

%% abbrvnat plainnat alpha amsplain unsrtnat
%%\bibliographystyle{unsrtnat}
% \bibliographystyle{D:/TEX/Lib/wbunsrtnat_short}
% \bibliography{D:/TEX/Lib/mybib}
%%\input main.bbl

\end{document}